\documentclass[reprint,pre,floatfix]{revtex4-1}

\usepackage{lineno,hyperref}
\modulolinenumbers[5]

\usepackage{todonotes}
\usepackage{makeidx}
\usepackage{graphics}
\usepackage{color}
\usepackage{listings}
\usepackage{bm}
\usepackage{url}
\usepackage{booktabs}
\usepackage{amsmath}
\usepackage{paralist}
\usepackage{amssymb}
\usepackage{amsthm}
\usepackage{subcaption}
\usepackage{algorithm}
\usepackage{algorithmic}

\graphicspath{ {img/} {plot/} }

\newcommand*\diff{\mathop{}\!\mathrm{d}}

\begin{document}



\title{Towards a unified lattice kinetic scheme for relativistic hydrodynamics}

\author{A. Gabbana}
\affiliation{Universit\`a di Ferrara and INFN-Ferrara, 
             Via Saragat 1, I-44122 Ferrara, Italy.}

\author{M. Mendoza}
\affiliation{ETH Z\"urich, Computational Physics for Engineering Materials, 
             Institute for Building Materials, Schafmattstra{\ss}e 6, HIF, 
             CH-8093 Z\"urich, Switzerland.}

\author{S. Succi}
\affiliation{Istituto per le Applicazioni del Calcolo C.N.R., 
             Via dei Taurini, 19 00185 Rome, Italy.}

\author{R. Tripiccione}
\affiliation{Universit\`a di Ferrara and INFN-Ferrara, 
             Via Saragat 1, I-44122 Ferrara, Italy.~}

\begin{abstract}
  We present a systematic derivation of relativistic lattice kinetic equations for finite-mass particles, 
  reaching close to the zero-mass ultra-relativistic regime treated in the previous literature. 
  Starting from an expansion of the Maxwell-J\"uttner distribution on orthogonal polynomials,  
  we perform a Gauss-type quadrature procedure and discretize the relativistic Boltzmann equation 
  on space-filling Cartesian lattices. The model is validated through numerical comparison with standard
  tests and solvers in relativistic fluid dynamics such as Boltzmann approach multiparton scattering (BAMPS) and
  previous relativistic lattice Boltzmann models.
  This work provides a significant step towards the formulation of a unified relativistic lattice kinetic scheme, 
  covering both massive and near-massless particles regimes. 
\end{abstract}




\maketitle

\section{Introduction}\label{sec:intro}

Relativistic kinetic theory and relativistic fluid dynamics play an increasingly important role in several fields of 
modern physics, with applications stretching over widely different scales, ranging from a very rich phenomenology in 
the realm of astrophysics \cite{shore-1992} down to atomic scales (e.g., in the study of the electron properties of graphene in 
effective 2D systems \cite{muller-2008} or the phenomenology of exotic states of quantum matter, such as the recently discovered 
Weyl fermion pseudo-particles \cite{wan-2011}), further down to subnuclear scales, in the realm of quark-gluon plasmas \cite{ackermann-2001}.
This motivates the quest for powerful and efficient computational methods, able to accurately study fluid dynamics in the
relativistic regime and possibly also to seamlessly bridge the gap between relativistic and low-speed non-relativistic fluid regimes.
Over the years, lattice kinetic theory has been at the basis of the development of increasingly complex and accurate 
lattice Boltzmann methods (LBM), able to simulate many relevant physics problems, including e.g., high Reynolds turbulent 
regimes, transport in porous media, multi-phase flows and many others \cite{succi-2001, aidun-2010, succi-2015}. 
%
One key advantage of most LBM algorithms lies in their computer-friendly structure, that has allowed the development of several massively parallel 
HPC implementations 
\cite{tolke-2008, bernaschi-2009, mountrakis-2015, calore-2016}.

The last decade has witnessed several attempts to develop LBM capable of handling the relativistic regime. 
The first model was developed by Mendoza et al. \cite{mendoza-2010a, mendoza-2010b}, based on the Grad's moment matching technique.
Romatschke et al. \cite{romatschke-2011} developed a scheme for an ultra-relativistic gas based on the expansion on orthogonal 
polynomials of the Maxwell-J\"uttner distribution, following a procedure similar to the one used for non-relativistic LBM.
However, this model is not compatible with a Cartesian lattice, thus requiring interpolation to implement the streaming phase.
Li et al. \cite{li-2012} have extended the work of Mendoza et al. using a multi relaxation-time collision operator,  
which, by independently tuning shear and bulk viscosity, has allowed the use of a Cartesian lattice. 
However, this model is not able to recover the third order moments of the distribution. 
Mohseni et al. \cite{mohseni-2013} have shown that it is possible to avoid multi-time relaxation schemes, still
using a D3Q19 lattice and properly tuning the bulk viscosity for ultra-relativistic flows, so as to recover only the 
conservation of the momentum-energy tensor.
This is a reasonable approximation in the ultra-relativistic regime, where the first order moment plays a minor role,
but leaves open the problem of recovering higher order moments.
A further step was taken in \cite{mendoza-2013}, with a relativistic lattice Boltzmann method (RLBM) able to recover higher 
order moments on a Cartesian lattice. This model provides an efficient tool for simulations in the ultra-relativistic regime.

All these developments use pseudo-particles of zero proper mass $m$ (or, more accurately, pseudo-particles for which the 
ratio particle mass over temperature, $m/T$, goes to zero). This implies that the equation-of-state appropriate for the 
fluid is the ultra-relativistic one, 
$\epsilon = 3 nT$, where $\epsilon$ is the energy density and $n$ the particle density. On the other hand, with the aim
of extending the range  of physical applications, one would like to explore wider ranges of the $m/T$ ratio 
and consider mildly relativistic, as well as ultra-relativistic regimes. 
From the algorithmic point of view, this discussion translates into the aim to develop a {\em 
unified} LBM, with the conceptual and technical ability to bridge the gap between
the ultra-relativistic regime ($u/c = \beta \simeq 1$, where $u$ is the fluid speed and $c$ the speed of light), 
all the way down to the non-relativistic one ($\beta \rightarrow 0$).

This work describes an initial step along this direction, introducing a new RLBM able to cover a
wider range of fluid velocities. In the development of the model, we follow a
procedure similar to the  one used  for many non-relativistic LBMs;
starting from an expansion of the Maxwell-J\"uttner distribution on orthogonal
polynomials,  we perform a Gauss-type quadrature procedure and
discretize the relativistic Boltzmann equation on space-filling Cartesian
lattices. We validate this RLBM by comparing with standard
tests and solvers in relativistic fluid dynamics and then present a
few simulation examples in the direction of prospective applications in astro and subnuclear physics.
Realistic applications, as well as hard-core computational aspects, are left for future work. 

This paper is structured as follows: In \autoref{sec:model}, we review the relativistic Boltzmann equation and present an overview
of the algorithmic steps involved in the development of our method. In \autoref{sec:discretization}, we describe in full detail 
the procedure used to discretize the relativistic Boltzmann equation on a Cartesian lattice.
In \autoref{sec:validation}, we numerically validate the model against some well-known relativistic flows, while in 
\autoref{sec:results} we present preliminary prospects of future physics applications. 
The paper ends with \autoref{sec:conclusions}, summarizing our results and future directions of research. 
Since the mathematics becomes quickly very involved,  many details are moved to appendices, while the most complex 
mathematical expressions are made available in the form of Supplemental Information \cite{supplementary-material}.

\section{Model Description}\label{sec:model}
In this section, we introduce the relativistic Boltzmann equation, and summarize our approach to its discretization in terms of a new RLBM; full details follow in the following section, so a self-contained description of our approach stretches across those two sections.

We  consider a single non-degenerate relativistic fluid whose quantum effects are not taken into account. 
The system is made up of particles with rest mass $m$; in kinetic theory, one is interested in the probability of finding a 
particle with momentum $\bm{p}$ at a given time $t$ and position $\bm{x}$; we adopt the usual relativistic notation,
$x^{\alpha} = \left( ct, \bm{x} \right)$ and  
$p^{\alpha} = \left( p^0, \bm{p} \right)$, with $\bm{x}$ and $\bm{p} \in \mathbb{R}^3 $.
The particle distribution function $f(\bm{x}, \bm{p}, t) = f(x^{\alpha}, p^{\beta})$ obeys the relativistic Boltzmann equation 
that, in the absence of external forces, reads:
\begin{equation}\label{eq:relativistic-boltzmann}
  p^{\alpha} \frac{\partial f}{\partial x^{\alpha}} = \Omega(f) \quad ,
\end{equation}
with an appropriate collision term $\Omega(f)$.
In the non-relativistic regime one usually replaces the collision term with the BGK approximation \cite{
bhatnagar-1954}; we adopt the relativistic generalization provided by the Anderson-Witting model \cite{anderson-witting-1974a, 
anderson-witting-1974b}: 
\begin{equation}\label{eq:anderson-witting}
  \Omega(f) = - \frac{U^{\alpha} p_{\alpha}}{\tau_{f} c^2} \left( f - f^{eq} \right) \quad ,
\end{equation}
with $\tau_{f}$ the relaxation (proper-)time, 
$U^{\alpha} = \gamma \cdot (c, \bm{u}) $ ($\gamma = 1 / \sqrt{ 1 - u^2 / c^2}$) 
the macroscopic four-velocity, and $f^{eq}$ the local equilibrium distribution, 
namely the Maxwell-J\"uttner distribution:
\begin{equation}\label{eq:maxwell-juttner}
  f^{eq} = \frac{1}{\mathcal{N}} \exp{ \left( - \frac{p_{\mu} U^{\mu}}{k_B T} \right) } \quad ;
\end{equation}
$\mathcal{N}$ is a normalization constant and $k_B$ the Boltzmann constant.
In the remainder of this paper we adopt units such that $c = 1$, $k_B = 1$.

Following Grad's theory \cite{grad-1949} the macroscopic description of a relativistic fluid is based on the moments of the
distribution function. We consider the first three moments of the distribution, namely the particle four-flow 
$N^{\alpha}$, the energy-momentum tensor $T^{\alpha \beta}$ and the third-order momentum $T^{\alpha \beta \gamma}$:
\begin{align}
  N^{\alpha             } &= \int f p^{\alpha}                      \frac{\diff \bm{p}}{p_0} \label{eq:mj-first-moment} \quad ,  \\
  T^{\alpha \beta       } &= \int f p^{\alpha} p^{\beta}            \frac{\diff \bm{p}}{p_0} \label{eq:mj-second-moment}\quad ,  \\
  T^{\alpha \beta \gamma} &= \int f p^{\alpha} p^{\beta} p^{\gamma} \frac{\diff \bm{p}}{p_0} \label{eq:mj-third-moment} \quad .
\end{align}

Hereafter we will use the subscript $E$ to refer to these tensors taken at the equilibrium, i.e. using $f^{eq}$ in place of $f$ in their definition.
It can be shown that, see e.g. \cite{cercignani-2002}, $N_E^{\alpha}$ and $T_E^{\alpha \beta}$ are given by
\begin{align}
  N_E^{\alpha       } &= n U^{\alpha}                                                             \label{eq:quadri-flux} \quad , \\
  T_E^{\alpha \beta } &= \left( \epsilon + P \right) U^{\alpha} U^{\beta} - P \eta^{\alpha \beta} \label{eq:energy-tensor};
\end{align}
$n$ is the particle number-density, $\epsilon$ the energy density, $P$ the pressure and $\eta^{\alpha \beta}$ the 
Minkowski metric tensor (that we write as $\eta^{\alpha \beta} = diag(1, -1, -1, -1) )$. 

The Anderson-Witting model correctly reproduces the conservation equations:
\begin{align}
  \partial_{\alpha} N^{\alpha}              &= 0 \quad , \\
  \partial_{\beta } T^{\alpha \beta}        &= 0 \quad . 
\end{align}

\subsection{Discrete relativistic Boltzmann equation}

We now describe our approach to derive a relativistic lattice Boltzmann equation, following a procedure similar to the one used with 
non-relativistic \cite{higuera-1989, xiaoyi-luo-1997, shan-1997, martys-1998} and earlier ultra-relativistic LBMs 
\cite{romatschke-2011, mendoza-2013}. We perform the following steps:
\begin{enumerate}
\item 
Write \autoref{eq:relativistic-boltzmann} in terms of quantities that can be discretized on a regular lattice. 
Following a standard procedure, we write the explicit expression of the relativistic lattice Boltzmann equation,
\begin{equation}\label{eq_boltz_mmj}
  p^0 \partial_t f + p^l \nabla_l f =  
  - \frac{p_{\mu} U^{\mu}}{\tau_{f}} (f - f^{eq}) \quad,       
\end{equation}
and divide left and right hand sides by $p^0$, obtaining:
\begin{equation} \label{eq:relativistic-boltzmann-v2}
  \partial_t f + v^l \nabla_l f = 
- \frac{p_{\mu} U^{\mu}}{\tau_{f} p^0} (f - f^{eq}) \quad,
\end{equation}
with $v^l = p^l / p^0$ the components of the microscopic velocity; 
in other words, we cast the equation in a form in which the time-derivative and the propagation term 
are the same as in the non-relativistic regime; the price to pay is an additional dependence on $p^0$ of the relaxation term.
\item 
Expand $f^{eq}$ in an orthogonal basis; we adopt Cartesian coordinates and use a basis of polynomials orthonormal with respect 
to a weight given by the Maxwell-J\"uttner distribution in the fluid rest frame ($\bm{u} = 0$):
\begin{equation}
  \omega(p^0) = \frac{1}{\mathcal{N}_R} \exp{\left( -p^0 / T \right)}.
\end{equation}
where, again, $\mathcal{N}_R$ is a normalization factor.
Call  $\{J^{(i)}, i = 1,2\dots \}$ these polynomials (that we compute in the following section); then 
\begin{equation}\label{eq:mj-projected-distribution}
  f^{eq}(\bm{p},U^{\mu}, T) = \omega( p^0) \sum_{k = 0}^{\infty} a^{(k)}(U^{\mu}, T) J^{(k)} (\bm{p}) , 
\end{equation}
where the projection coefficients $a^{(k)}$ are
\begin{equation}\label{eq:mj-projection-coefficients}
  a^{(k)}(U^{\mu}, T) = \int f^{eq}(\bm{p},U^{\mu}, T)  J^{(k)}(\bm{p}) \frac{\diff \bm{p}}{p^0}. 
\end{equation}
The approximation $f^{N eq}(\bm{p}, U^{\mu}, T )$, obtained truncating the summation in  \autoref{eq:mj-projected-distribution} 
to the $N$-th order, recovers the same moments as the original distribution function to order $N$, given that the 
expansion coefficients correspond to the moments of the distribution. For example, a third order expansion ensures that the results 
of the integrals in \autoref{eq:mj-first-moment}, \ref{eq:mj-second-moment} and \ref{eq:mj-third-moment} are correctly recovered.
\item
Find a Gauss-like quadrature on a regular Cartesian grid able to reproduce correctly the moments of the original distribution up 
to order $N$. We proceed in such a way as to preserve one of the most important features of lattice Boltzmann models, namely exact 
streaming; this means that all quadrature points $v^l_i = p^l_i/p^0$ must sit on lattice sites.
At this point, the discrete version of the equilibrium function reads as follows:
\begin{equation}\label{eq:mj-discrete-feq}
  f_i^{N eq} = w_i \sum_{k = 0}^{K_N} a^{(k)}(U^{\mu}, T) J^{(k)} (p^{\mu}_i) \quad ,
\end{equation}
with $w_i$ appropriate weights, and $K_N$ is the number of orthogonal polynomials up to the order $N$.
\item
Use the above result to write the discrete relativistic Boltzmann equation:
\begin{equation}\label{eq:discrete-rlb-andersonwitting}
  f_i(\bm{x} + v^{i} \Delta t, t + \Delta t) - f_i(\bm{x}, t) = - \Delta t~ \frac{p_i^{\mu} U_{\mu}}{p^0 \tau} (f_i - f_i^{eq}),
\end{equation}
where $v_{i}$ are the microscopic lattice velocities of each streaming population, and $\tau$ the relaxation time in
lattice units (whose relation with $\tau_{f}$ will be discussed in \autoref{sec:validation})
\end{enumerate}

\autoref{eq:discrete-rlb-andersonwitting} allows to simulate the evolution of the system in discrete space and time.
Once the $f_i$ are known, one computes the energy-momentum tensor (\autoref{eq:mj-second-moment}) as:
\begin{equation}\label{eq:mj-discrete-second-moment}
  T^{\alpha \beta       } =   \sum_i f_i p^{\alpha}_i p^{\beta}_i \quad .
\end{equation} 

The Anderson-Witting collisional model is only compatible with the 
Landau-Lifshitz decomposition \cite{cercignani-2002}, which implies
\begin{align}
  n                   =&~   U_{\alpha} N^{\alpha}      \label{eq:landau-lifshitz-first-moment } \quad ,  \\
  \epsilon U^{\alpha} =&~ T^{\alpha \beta } U_{\alpha} \label{eq:landau-lifshitz-second-moment} \quad ,
\end{align}
%
so we obtain the energy density $\epsilon$ and $U_{\alpha}$ 
solving the eigenvalue problem in \autoref{eq:landau-lifshitz-second-moment}.
Finally temperature is linked to energy and particle density via a suitable equation of state. 

Note that \autoref{eq:landau-lifshitz-first-moment } and \ref{eq:landau-lifshitz-second-moment} stem from the 
property of the collision operator to conserve the number of particles and their energy. 
As a result, its zeroth and first order moments are bound to vanish. 
Thus, for instance, in the continuum case, right hand side of \autoref{eq_boltz_mmj}, we calculate the respective 
moments of the collision operator:
\begin{equation}
  - \int \frac{U_\mu}{\tau_f} (f p^\mu - f^{eq} p^\mu) \frac{\diff \bm{p}}{p^0} \quad = \frac{1}{\tau_f} (U_\mu N^\mu -  U_\mu N_E^{\mu} ), 
\end{equation}
\begin{equation}
  - \int \frac{U_\mu}{\tau_f} (f p^\mu p^\nu - f^{eq} p^\mu p^\nu) \frac{\diff \bm{p}}{p^0} = \frac{1}{\tau_f} (U_{\mu} T^{\mu \nu} -  U_\mu T_E^{\mu \nu} ) \quad , 
\end{equation}
and due to the fact that these two expression should be equal to zero, we get 
\begin{equation}
  U_\mu N^\mu =  U_\mu N_E^{\mu} \quad, 
\end{equation}
\begin{equation}
  U_{\mu} T^{\mu \nu}  = U_\mu T_E^{\mu \nu} \quad . 
\end{equation}
Since we know the equilibrium moments, it can be shown that
\begin{equation}
  U_\mu N^\mu =  U_\mu N_E^{\mu} = n\quad, 
\end{equation}
\begin{equation}
  U_{\mu} T^{\mu \nu}  = U_\mu T_E^{\mu \nu} = \epsilon U^\nu \quad . 
\end{equation}
However, it is important to observe that these expressions do not imply that $N^\mu = N_E^\mu$ and $T^{\mu \nu} = T_E^{\mu \nu}$, 
but rather that the non-equilibrium components of $N^\mu$ and $T^{\mu \nu}$ are orthogonal to the four-velocity. 
The same is true in the discrete case, with integrals replaced by summations over the set of discrete velocities.

\subsection{Equation of State}
As outlined in the introduction, we consider the case of (in-principle) arbitrary values of the particle mass 
(and hence of the $m/T$ ratio); this allows to consider general equations of state (EOS) not
confined to the ultra-relativistic limit;

In the ultra-relativistic regime the EOS is well known:
\begin{equation}\label{eq:ultrarelativistic-eos}
  \epsilon = 3 n T \quad ,
\end{equation}
A more general EOS for a perfect gas -- valid for any value of the $m/T$ ratio -- has been derived several decades ago by Karsch et 
al. \cite{karsch-1980}: 
\begin{align}\label{eq:relativistic-eos}
  \epsilon - 3 n T &= (n T)~\frac{m}{T} \frac{K_{1}(m / T)}{K_{2}(m / T)} \quad ,\\
  P &= n~T \quad ;
\end{align}
here and in the following, $K_i$ is a modified Bessel function of the second kind of index $i$.
Note that $x K_1(x)/K_2(x) \rightarrow 0$ as $x \rightarrow 0$, so \autoref{eq:relativistic-eos} correctly reproduces the 
ultra-relativistic limit (\autoref{eq:ultrarelativistic-eos}) as $m/T \rightarrow 0$.
For the non-relativistic limit, one writes
\begin{equation}\label{eq:nonRel1}
  (\epsilon - n~m)- 3 n T = 
  (n T)~(\frac{m}{T} \frac{K_{1}(m / T)}{K_{2}(m / T)} - m/T) \quad .
\end{equation}
Noting that $(x~K_1(x)/K_2(x) -x) \rightarrow - 3/2$ as $x \rightarrow \infty$, and defining $\epsilon_c = \epsilon - n~m$ 
(the non-relativistic kinetic energy density), we also recover the well-known non-relativistic expression $\epsilon_c = 3/2~nT$. 
It is also interesting to look at the difference between \autoref{eq:relativistic-eos} and \autoref{eq:ultrarelativistic-eos} 
in the intermediate regimes.
To this effect, we 
rearrange \autoref{eq:relativistic-eos} in the following way:
\begin{equation}\label{eq:eos-difference}
  \frac{\epsilon}{3 n T}  =   1 + \frac{1}{3} \frac{m}{T} ~ \frac{K_{1}\left( \frac{m}{T} \right) }{K_{2}\left( \frac{m}{T} \right)},
\end{equation} 
explicitly highlighting the ratio between the two EOS. 

\begin{figure}[t]
  \centering
  \includegraphics[width=.49\textwidth]{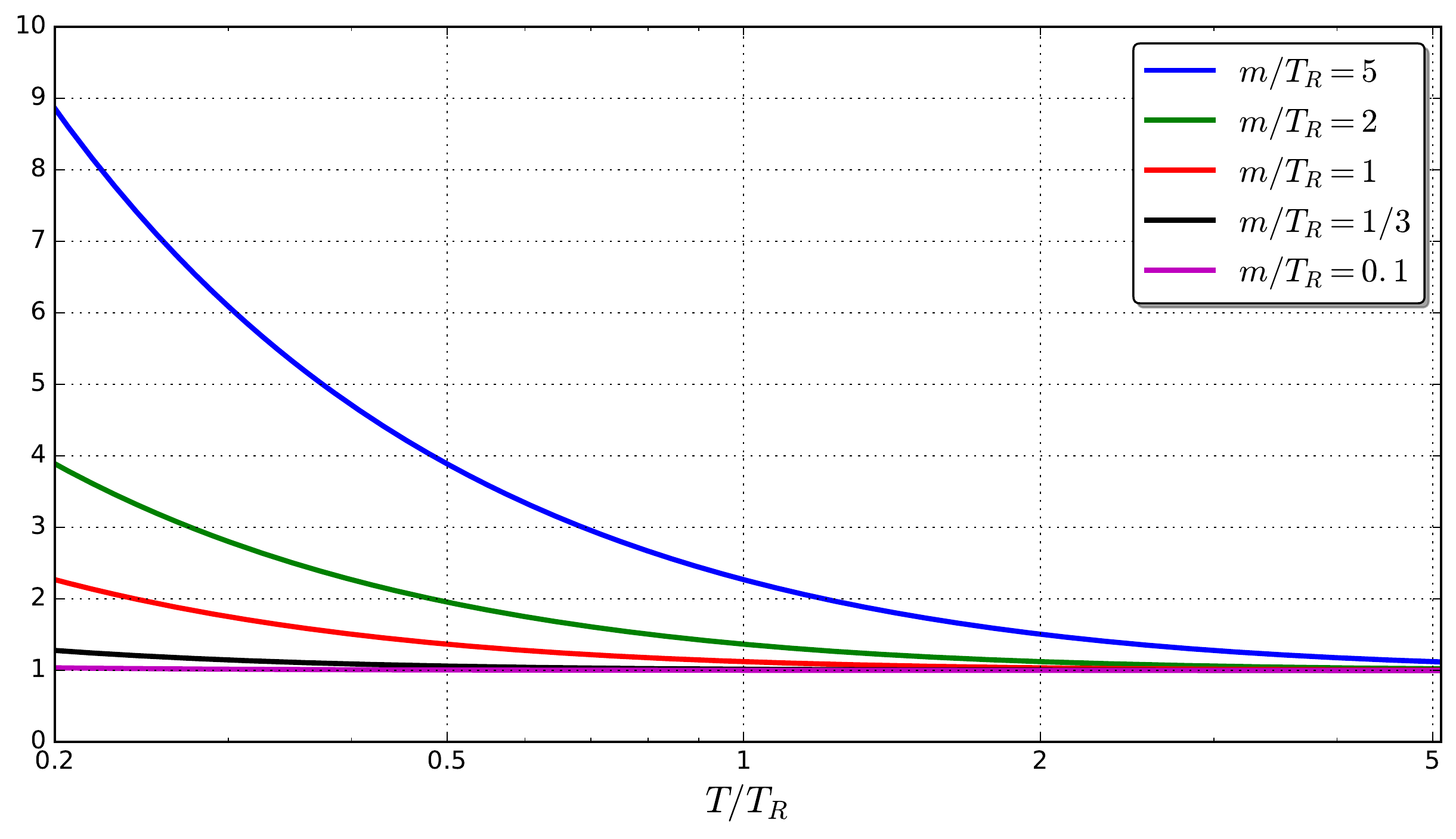}
  
  \caption{Plot of the RHS of \autoref{eq:eos-difference} as a function of $T$, for different values of $m$,
           in arbitrary units.
          }\label{fig:edensity-ration}
\end{figure}

This quantity is plotted in \autoref{fig:edensity-ration} as a function of $T$ for selected values of $m$.

\subsection{Transport Coefficients}

The transport coefficients of the model, i.e. shear and bulk viscosities and thermal conductivity, are defined as usual from the non-equilibrium
contributions of the energy-momentum tensor \cite{cercignani-2002}. The shear viscosity can be obtained by using the following expression,
\begin{equation}
    2\eta ~ \partial^{<\alpha} U^{\beta >} = \left( \Delta_\gamma^\alpha \Delta_\delta^\beta - \frac{1}{3} \Delta^{\alpha \beta} \Delta_{\gamma \delta}   \right) T^{\gamma \delta} \quad ,
\end{equation}
where $\Delta^{\alpha \beta} \equiv \eta^{\alpha \beta} - U^\alpha U^\beta$, and the expression $\partial^{<\alpha} U^{\beta >}$ stands for
\begin{equation}
    \partial^{<\alpha} U^{\beta >} = \left [\frac{1}{2}\left( \Delta_\gamma^\alpha \Delta_\delta^\beta + \Delta_\delta^{\alpha} \Delta_{\gamma}^{\beta} \right)      - \frac{1}{3} \Delta^{\alpha \beta} \Delta_{\gamma \delta} \right ] \partial^\gamma U^\delta \quad .
\end{equation}
The bulk viscosity $\kappa$, on the other hand, can be calculated by using
\begin{equation}
 - \kappa ~ \partial_\alpha U ^\alpha  = -P - \frac{1}{3} \Delta_{\alpha \beta} T^{\alpha \beta} \quad ,
\end{equation}
and finally, the thermal conductivity $\lambda$, with the expression
\begin{equation}
  \lambda \left ( \partial^\alpha T - T U^\beta \partial_\beta U^\alpha \right ) = \Delta_{\gamma}^{\alpha} U_\beta T^{\beta \gamma} \quad .
\end{equation}

It is important to mention that, unlike the non-relativistic case, there is no straightforward way to compute the transport 
coefficients directly from the model parameters, since it is known that the Chapman-Enskog and the Grad procedure deliver
(slightly) different results \cite{cercignani-2002}. 
There are also other kind of expansions developed for that purpose \cite{israel-1976, denicol-2012} and yet an unique expression has not been found. 
Following Mendoza et al. \cite{mendoza-2013}, in this work we shall assume the transport coefficients delivered by the Grad procedure.
See later for further discussions on this point.

\section{Lattice Discretization}\label{sec:discretization}

In this section we describe in details all steps, outlined in the previous section, 
required to implement a relativistic lattice Boltzmann procedure, that is, 
all the ingredients necessary to define and evolve \autoref{eq:discrete-rlb-andersonwitting}.

\subsection{Relativistic orthonormal polynomials}\label{sec:relativistic-poly}
We start by constructing an orthonormal basis of polynomials.
Following Mendoza et al. \cite{mendoza-2013} we adopt the 
equilibrium distribution in the co-moving frame as our weight function:
\begin{equation}\label{eq:mj-poly-weighting-factor}
  \omega(p^0) = \frac{1}{\mathcal{N}_R} \exp{\left( -p^0 / T_R \right)} \quad .
\end{equation}
%
Hereafter we will use $T_R$ as a normalization factor to write adimensional quantities 
and to convert from physics to lattice units.

In order to construct a set of orthogonal polynomials we apply the well known Gram-Schmidt procedure, starting from the set
$ \mathcal{V} = \{ 1, p^{\alpha}, p^{\alpha} p^{\beta}, \dots \}$.
To carry out this procedure, one must compute integrals of the
form
\begin{equation}\label{eq:mj-integrals}
   I^{\alpha \beta \gamma \dots} =  
   \int \exp{ \left( - \frac{ p_{\mu} U^{\mu} }{ T } \right) } p^{\alpha} p^{\beta} p^{\gamma} \dots  \frac{\diff^3 p}{p_0} \quad ,
\end{equation}
that can be written in terms of Bessel functions\cite{cercignani-2002}.
For example: 
\begin{equation}
  I = \int \exp{ \left( - \frac{ p_{\mu} U^{\mu} }{ T } \right) } \frac{\diff^3 p}{p^0}
  =
  4 \pi ~\frac{m}{T} ~T^2 K_1 \left(  \frac{m }{T}  \right) \quad ,
\end{equation}
\begin{equation}
  I^{\alpha} = \int \exp{ \left( - \frac{ p_{\mu} U^{\mu} }{ T } \right) } p^{\alpha} \frac{\diff^3 p}{p^0}
  =
  4 \pi ~(\frac{m}{T})^2 ~T^3 K_2 \left(  \frac{m }{T}  \right) U^{\alpha} ;
\end{equation}
integrals with higher powers of $p$ are derived by differentiating with respect to $m/T$ and taking into 
account well-known properties of the Bessel functions.

It is useful to normalize  $\omega(p^0)$ so that
\begin{equation}
  \int \omega(p^0) \frac{\diff^3 p}{p_0} = 1 \quad ,
\end{equation}
implying that 
\begin{equation}\label{eq:cercignani-recurrence-relation}
  \mathcal{N}_R = 4 \pi \frac{m}{T_R} T_R^2 K_1 \left(  \frac{m }{ T_R}  \right) = 4 \pi \bar{m} T_R^2 K_1(\bar{m}) \quad ,
\end{equation}
where we adopt the shorthand $\bar{m} = m/T_R$. 

The complete set of polynomials up to the $2$-nd order has $14$ independent elements while $30$ elements are needed 
at the $3$-rd order.
See Appendix \ref{sec:appendixB} for all
$2$-nd order polynomials, while the complete set 
(up to $3$-rd order) is available as supplemental material; we label all polynomials as $J^{(n)}_{k_1 \cdots k_n}$, 
where $n$ is the order of the polynomial and the $k$ indexes corresponds to the components of $p^{\mu}$ they depend upon.
As an example, the first non-constant polynomial is
\begin{equation}
  J_0^{(1)} =   \frac{1}{ A } \left(  \frac{p_0}{T_R} - \bar{m} \frac{  K_2(\bar{m}) }{ K_1(\bar{m}) }   \right)
 = \frac{1}{ A } \left(  \bar{p}_0 - \bar{m} \frac{  K_2(\bar{m}) }{ K_1(\bar{m}) }   \right),
\end{equation}
All polynomials are adimensional so we write them in terms of $\bar{p}^{\alpha} = p^{\alpha} / T_R$; all coefficients, including the normalization constant $A$, only depend on $\bar{m}$.

We now compute  $a^{(k)}(U^{\mu}, T)$, the projections of $f^{eq}(p,U^{\mu},T)$ (see \autoref{eq:mj-projection-coefficients}), 
in a generic reference frame. We choose to normalize $f^{eq}$ so that
\begin{equation}
  n U^{\alpha} = N_E^{\alpha} = \int f^{eq} p^{\alpha} \frac{\diff \bm{p}}{p_0} \quad ,
\end{equation}
implying 
\begin{equation}\label{eq:maxwell-juttner-normalization}
  \mathcal{N} =  4 \pi \left( \frac{m}{T} \right)^2 T^3 K_2 \left(  \frac{m }{T}  \right) .
\end{equation}
The computation of these coefficients  is a tedious but straightforward task, 
as it implies again integrals of the form of \autoref{eq:mj-integrals}. 
The coefficients of all polynomials up to the $2$-nd order are listed in Appendix \ref{sec:appendixC} and all remaining ones are
available as  supplemental material; coefficient labeling follows the same rules as for polynomials. 
For example, the explicit expression of $a_0^{(1)}$ reads:
\begin{align*}
  a_0^{(1)} & = \int f^{eq}  J_0^{(1)} \frac{\diff p^3}{p^0}  \\
            & = \frac{1}{\mathcal{N}} \frac{1}{A} \int \exp{ \left( -\frac{ p_{\mu} U^{\mu} }{ T } \right) } \left[ \frac{p_0}{T_R} - \bar{m} \frac{K_2(\bar{m})}{K_1(\bar{m})} \right] \frac{\diff p^3}{p^0} \\
            & = \frac{1}{T_R} \frac{1}{A} \left( U_0 - \frac{ K_2(\bar{m})}{K_1(\bar{m})} \frac{ K_1(\frac{m}{T})}{K_2(\frac{m}{T})} \right) \quad .     
\end{align*}
All projection coefficients $a^{(k)}$ carry a dimension of one over temperature (or energy); correspondingly, 
we write them as an explicit $1/T_R$ prefactor followed by adimesional expressions written in terms of $m/T_R = \bar{m}$ and $m/T$.

\subsection{Polynomial expansion of the distribution function at equilibrium}
\begin{figure}[t]
\centering
\includegraphics[width=0.49\textwidth]{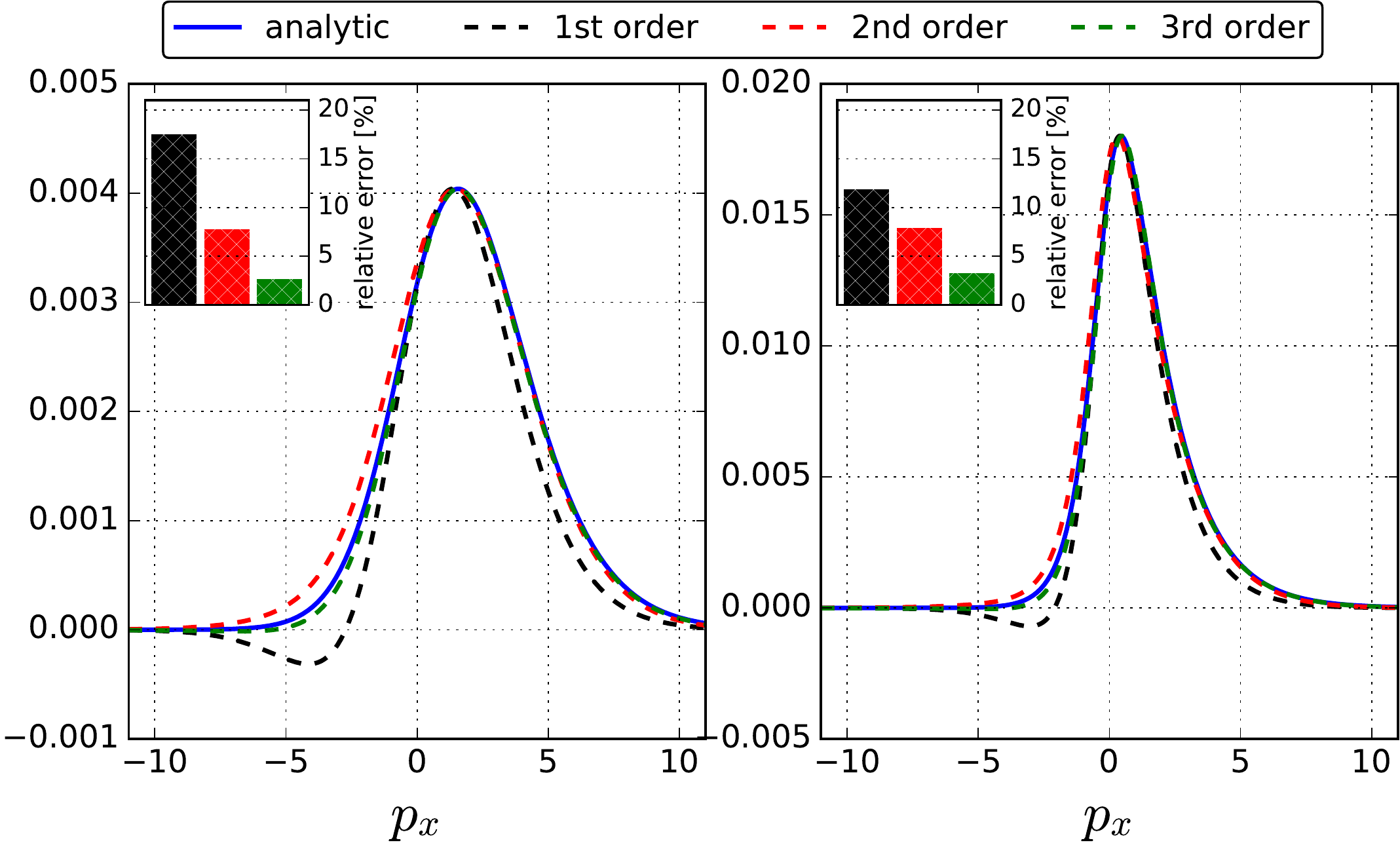}
\caption{Comparison of the analytic Maxwell J\"uttner distribution against first, 
        second and third order approximations computed using an orthogonal polynomial basis.
        Left:  $m = 5$, $T = 1$, $\bm{p} = (p_x, 0, 0)$ and $\bm{u} = (0.3, 0, 0)$;
        Right: $m = 1$, $T = 1$, $\bm{p} = (p_x, 0, 0)$ and $\bm{u} = (0.4, 0, 0)$.
        For each plot, column bars represent the percentage relative L2-error of each approximation with
        respect to the analytic distribution.
}\label{fig:mj-approx}
\end{figure}

It is now possible to write a polynomial approximation to $f^{eq}$ at any order $N$, via
\autoref{eq:mj-projected-distribution}, using the explicit expressions for the polynomials and for the expansion 
coefficients computed in the previous subsections. 
The analytic expressions quickly become very awkward; for example, the expansion of $f^{eq}$ at first order has $5$ 
terms and reads:
\begin{equation}
\begin{split}
& f^{eq}(p,U,\frac{m}{T_r},\frac{m}{T})  = \omega(p^0)  \frac{1}{T_R}  \left(   \frac{1}{A^2} \left(\bar{p}_0 - 
                  m \frac{ K_2\left(\bar{m}\right)}{K_1\left( \bar{m} \right)}\right) \right.\\
                  & \left(U_0-\frac{K_1\left(\frac{m}{T}\right) 
                  K_2\left(\bar{m}\right)}{K_2\left(\frac{m}{T}\right) K_1\left(\bar{m}\right)}\right) 
                  +  \frac{1}{B^2} \left( \bar{p}_x U_x + \bar{p}_y U_y + 
                  \bar{p}_z U_z \right) \\
                  & \left. +  \frac{K_1\left(\frac{m}{T}\right)}{\bar{m} K_2\left(\frac{m}{T}\right)} \right) \quad ,
\end{split}
\end{equation}
with the coefficients $A$ and $B$ defined in Appendix \ref{sec:appendixA}.
In general, upon factoring out the term $1/T_R$, the expression of $f^{eq}$ only depends on the ratios 
$m / T_R = \bar{m}$ and $T / T_R$  (in fact we can always write $\frac{m}{T} = \frac{m}{T_R} \frac{T_R}{T}$). As we
will see later, $\bar{m}$ is fixed by the quadrature, while $T/T_R$ controls the translation from physical to lattice units.
In \autoref{fig:mj-approx} we compare approximations at the first, second and third order against
the analytic Maxwell J\"uttner distribution, for several values of $m$, $T$, and $\bm{u}$;
as expected, the first order approximation fails to reproduce the analytic behavior, while the 
second and third order expansions provide increasingly accurate approximations.
%
\subsection{Quadratures with prescribed abscissa}\label{sec:quadrature}
In order to implement an RLBM on a Cartesian space-filling lattice 
we need to find the weights and the abscissas of a quadrature satisfying the
following orthonormal condition:
%
%
\begin{equation}\label{eq:ortho1}
\begin{split}
  \int \omega(\bar{p}^0) J^{(l)}(\bar{p}^{\mu}) J^{(k)}(\bar{p}^{\mu}) \frac{d^3 \bar{p}}{\bar{p}^0} 
  &= 
  \sum_n w_n~J^{(l)}(\bar{p}^{\mu}_n)J^{(k)}(\bar{p}^{\mu}_n) \\
  &= \delta_{lk} \quad ,
\end{split}
\end{equation}
where $\{J^{(i)}, i = 1,2\dots K\}$ are the orthogonal polynomials derived in \autoref{sec:relativistic-poly}, 
$p^\mu_n$ are the four-momentum vectors defined at appropriate points
in momentum space and the $w_n$ are suitable weights \cite{philippi-2006}. Our goals is to satisfy the above equation up to the sixth 
order in $p$, so up to the fifth order of the equilibrium distribution is recovered. 

As already discussed, we want to ensure exact streaming, that is we require that all $p^\mu_n$ sit exactly on sites 
of our Cartesian grid.
We can fulfill this requirement, since we work with a finite value of the particle mass $m$ (hence with a finite value of $m/T_R$). 
To this effect, we adopt populations belonging to several particle groups $G$, each group defining (pseudo-)particles that, at each 
time step, move from one lattice site to other sites at a given fixed distance. A large list of groups that we can select from is 
collected in Appendix \ref{sec:appendixC}. For instance, the well-known non-relativistic D3Q19 model uses the set $\{G_1,G_2,G_3\}$. 
Consequently, \autoref{eq:ortho1} becomes
%
\begin{equation}\label{eq:orthonormal}
\begin{split}
  \int \omega(\bar{p}^0) J^{(l)}(\bar{p}^{\mu}) J^{(k)}(\bar{p}^{\mu}) \frac{d^3 \bar{p}}{\bar{p}^0} 
       &= 
       \sum_i \sum_j w_j J^{(l)}(\bar{p}^{\mu}_{i,j})J^{(k)}(\bar{p}^{\mu}_{i,j}) \\
       &= \delta_{lk} \quad ,
\end{split}
\end{equation}
with $\bar{p}^{\mu}_{i,j}$ corresponding to the $i-$th element of the $j-$th group and $w_j$ is the weight of the $j$-th group.

\begin{figure}[t]
\begin{subfigure}{0.235\textwidth}
\includegraphics[width=\linewidth]{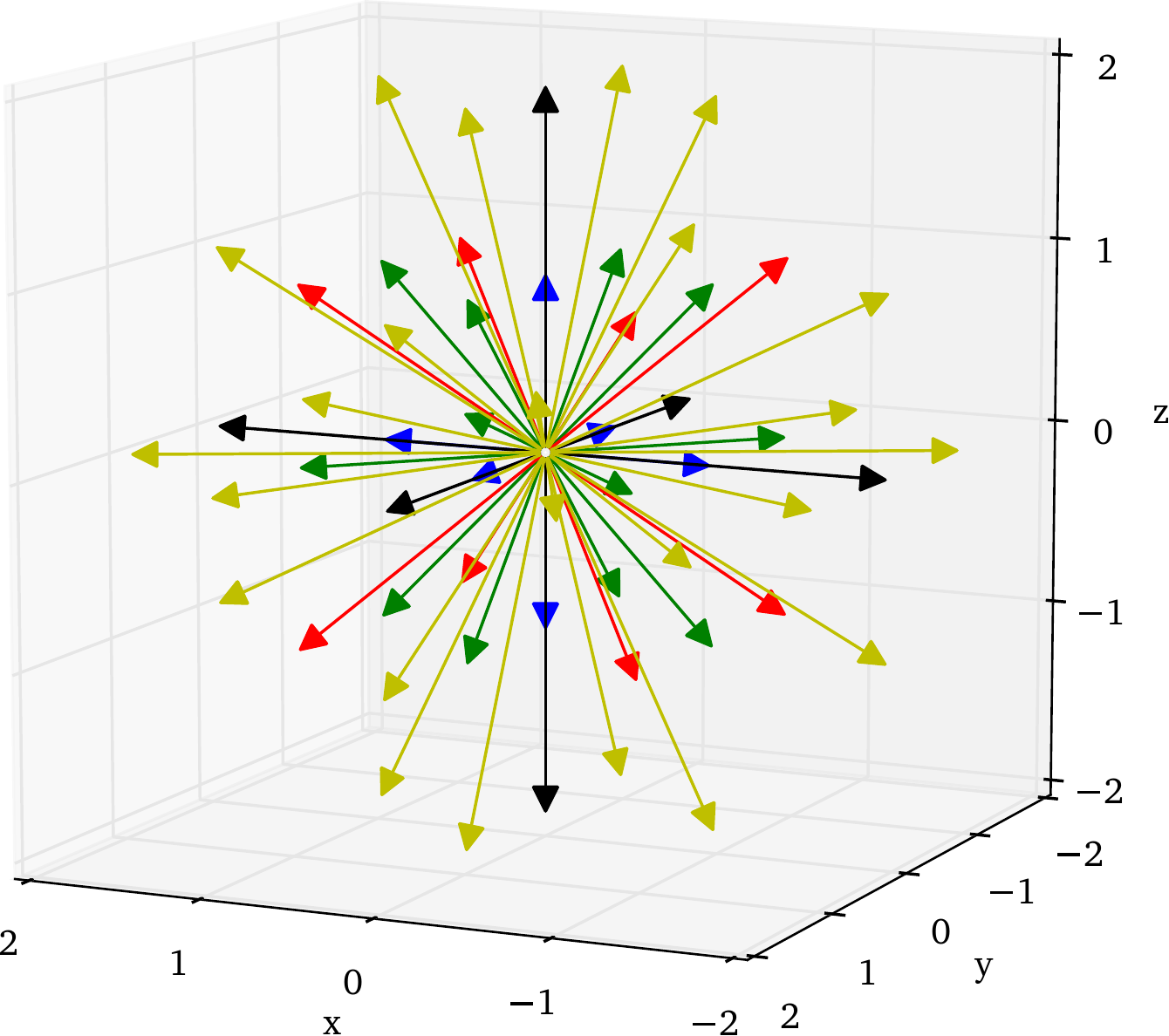}
\caption{} \label{fig:stencil-o2}
\end{subfigure}
\hspace*{\fill} 
\begin{subfigure}{0.235\textwidth}
\includegraphics[width=\linewidth]{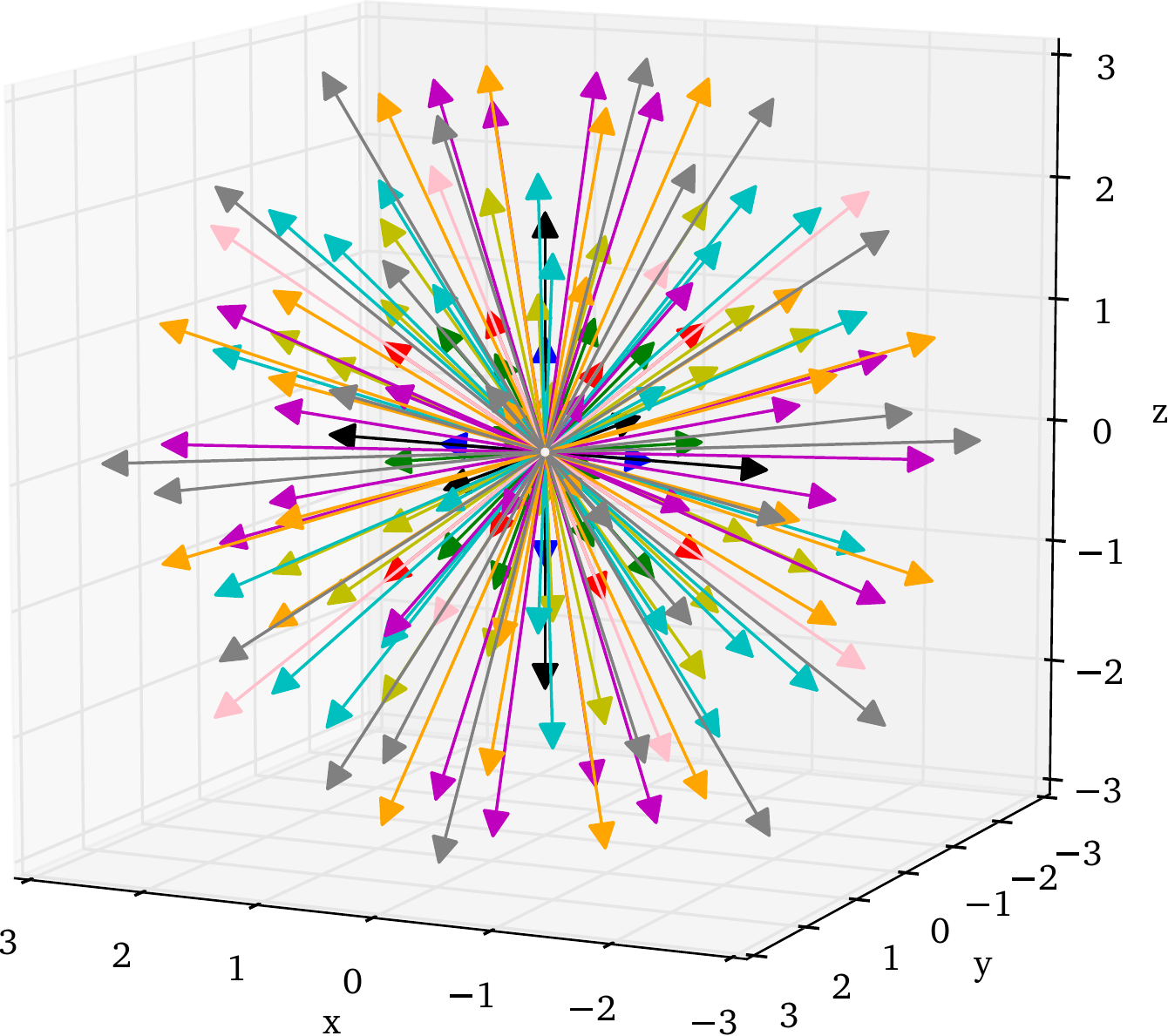}
\caption{} \label{fig:stencil-o3}
\end{subfigure}
\caption{Examples of stencils for the relativistic lattice Boltzmann method.
a) Stencil for a second order approximation formed by G = (G1, G2, G3, G4, G5, G6), with $57$  pseudo-populations;
b) stencil for a third order approximation formed by G = (G1, G2, G3, G4, G5, G8, G9, G10, G12, G13, G15), with $161$ pseudo-populations.
}\label{fig:stencil-examples}
\end{figure}

When using more than one group, we ensure exact streaming requiring
that velocities of particles belonging 
to each group are proportional to the (Cartesian-) distance they have to \emph{travel} to reach their destination. This means that 
different groups belong to different energy shells. Indeed, we write
\begin{equation}\label{eq:four-momentum}
p^\mu_{i,k} = m \gamma_k (1, v_0 \vec{n}_{i,k}) \quad .
\end{equation}
Here:
\begin{itemize}
  \item $ \vec{n}_{i,k} = (n^x_{i,k}, n^y_{i,k}, n^z_{i,k}) \in  \mathbb{N}^3$ are the coordinates of the $i-th$ element of 
         the $k-th$ group of the stencil; $|| \vec{n}_{k} || $ is the common value of $|| \vec{n}_{i,k} || $ for all vectors 
         belonging to group $k$.
  \item $\gamma_k$ is the relativistic $\gamma$ factor associated to $v_k = v_0~|| \vec{n}_k ||$ .
  \item $v_0$ is a common velocity parameter that can be freely chosen under the condition that $v_k \leq 1, \forall k$.
\end{itemize}
Otherwise stated, the set of the $\vec{n}_{i,k}$ defines the travel path of each element of each group; then, once $v_0$ has been 
set to a specific value, all elements of the group are assigned to an energy shell per \autoref{eq:four-momentum}. 
\autoref{fig:stencil-examples} shows examples of lattices compatibles with the requirements of \autoref{eq:orthonormal} 
and \autoref{eq:four-momentum}.

Assuming that a quadrature has been found and a suitable value for $v_0$ has been selected, 
\autoref{eq:discrete-rlb-andersonwitting} becomes

\begin{equation}\label{eq:discrete-relativistic-boltzmann}
  f_i(\vec{x} + v_0 \vec{n}_{i,k} \delta t, t + \delta t) - f_i(\vec{x},t) = 
 - \delta t~ \frac{p_i^{\mu} U_{\mu}}{p^0_i \tau} (f_i - f_i^{eq}) \quad .
\end{equation}

Therefore, our requirement 
\begin{equation}\label{eq:v0dt=dx}
  v_0 \vec{n}_{i,k}\delta t = \vec{N}_{i,k} \delta x \quad ,
\end{equation}
where $\delta x$ is the lattice spacing and $\vec{N}_{i,k}$ are integer numbers, is equivalent to a relation between  
time and space units on the lattice. 

\subsection{Finding quadratures}
Assuming for the moment that a certain set of particle groups has been selected, our
next step is to find the weights $w_j$ of a quadrature that solves
\autoref{eq:ortho1}, that we copy here for convenience, up to a prescribed order:
%
\begin{equation}\label{eq:ortho2}
\begin{split}
  \int \omega(\bar{p}^0) J^{(l)}(\bar{p}^{\mu}) J^{(k)}(\bar{p}^{\mu}) \frac{d^3 \bar{p}}{\bar{p}^0} 
  &= 
  \sum_n w_n~J^{(l)}(\bar{p}^{\mu}_n)J^{(k)}(\bar{p}^{\mu}_n) \\
  &= \delta_{lk} \quad ,
\end{split}
\end{equation}
where $p^\mu_{i,k}$ are four-momentum vectors defined in \autoref{eq:four-momentum}. Recall that the values of the $p^\mu_{i,k}$
depend on the group to which they belong and on a common hitherto arbitrary value for $v_0$.

We follow the procedure described in \cite{shan-2010}, 
building a lattice by adding as many groups as necessary to fulfill \autoref{eq:ortho2}.
For example, considering quadratures giving a second-order approximation, the system of \autoref{eq:orthonormal} has
$6$ linearly independent components , so one needs to build a stencil with (at least) $6$ different groups.
Likewise, at third order there are $11$ independent components, so we need $11$ groups. Yet higher order approximations
would require stencils with even larger numbers of groups.
\begin{figure}[t]
  \centering
  \includegraphics[width=.49\textwidth]{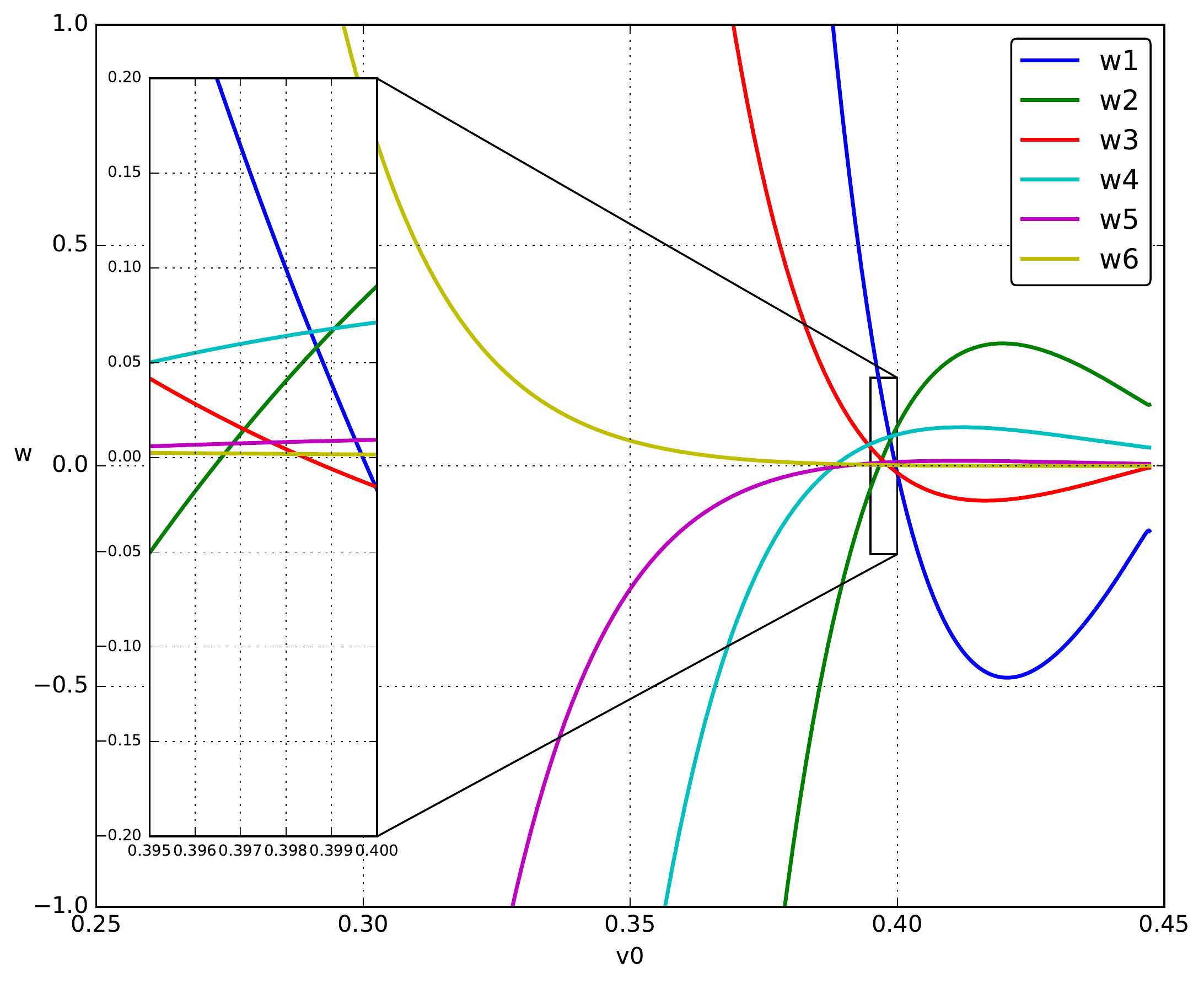}
  
  \caption{Parametric solution of the system of equations given by \autoref{eq:orthonormal} using the stencil 
           G = (G1, G2, G3, G4, G5, G6) with $\bar{m} = 5$. In this case we can identify a region for which 
           $w_i (v_0) \geq 0 ~ \forall i$  (orange coloured interval), giving a set of solutions that can be used 
           to build a numerically stable quadrature.
          }\label{fig:v0-vs-w}
\end{figure}
\autoref{eq:orthonormal} is a linear system in the unknowns $w_j$, whose coefficients in principle depend on $m/T_R$, on the 
chosen set of groups and on $v_0$. We look for solutions in which  $w_j \ge 0$ for all $j$s, as this improves numerical stability 
and is consistent with a (pseudo-)particle interpretation of the RLBM.  In practice, one: i) assigns a value for $m/T_R$; ii) 
selects a large enough set of particle groups; and then iii) solves \autoref{eq:orthonormal} for arbitrary values of $v_0$.

Let us see with an example at $2$-nd order the result of this procedure; we take $\bar{m} = 5$ and consider the stencil formed by 
the union of the first six groups in \autoref{tab:stencil-groups}: G = (G1, G2, G3, G4, G5, G6 ). With this stencil, the longest 
displacement is given by $G6$ having length $1 / \sqrt{5}$, 
so $v_0 \in [ 0, 1 / \sqrt{5} \approx 0.447 )$
as pseudo-particles cannot travel faster than light. \autoref{fig:v0-vs-w} shows the values of the $w_j$s 
that solve \autoref{eq:ortho2} as a function of $v_0$. We see that their values wildly oscillate between large positive 
and negative values; we can however identify a range of $v_0$ values ( $v_0 \in ( 0.3966 , 0.3984 )$), for which all weights are 
positive, providing acceptable solutions to the problem.

Taking for example $v_0 = 0.398$ the corresponding weights for the quadrature are:
\begin{align*}
  w_1 = 0.0993921725\dots ~~ w_2 = 0.0404025909\dots \\
  w_3 = 0.0043631818\dots ~~ w_4 = 0.0640885469\dots \\   
  w_5 = 0.0081185158\dots ~~ w_6 = 0.0018506095\dots
\end{align*}
%
%
Particularly useful values of $v_0$ are those located at the boundaries of the interval since, as easily seen in 
\autoref{fig:v0-vs-w}, in this case some weights become zero thus pruning certain lattice velocities.
In our example one can reduce the set of $57$ velocities to $51$ by setting $w_2$ to zero 
(taking $v_0 = 0.3965826549\dots$), or to $45$ by setting $w_3$ to zero ($v_0 = 0.3984063950\dots$).
More examples, and accurate values for the weights, are provided in the supplemental material.

In general, many different solutions to the quadrature problem exist. Indeed, one first has the freedom to 
arbitrarily choose the particle groups (that in turn define the corresponding set of momentum four-vectors
per \autoref{eq:four-momentum}) and the reference value for $\bar{m}$ and then one has to pick up a particular value for $v_0$.
From an algebraic point of view, \autoref{eq:ortho2} leads to a linear system
of equations, parametric on $v_0$:
\begin{equation}
  A(v_0) \bm{w} = \bm{b} \quad .
\end{equation}
Here $A$ is a $l \times k$ matrix ($l$ being the number of possible combinations of the orthogonal polynomials, $k$
the number of groups forming the stencil), $\bm{b}$ is a known binary vector, and $\bm{w}$ is the vector of unknowns.
Since the Gaussian quadrature requires strictly positive weights in order to guarantee numerical stability,
we need to select values of $v_0$ (if they exist) such that $w_i > 0 ~ \forall i$. For low-order approximations it is possible 
to compute an analytic solution, writing  each weight $w_i$ as an explicit function of the free parameter $v_0$, but this become 
quickly very hard and, already at the second-order, numerical solutions are necessary.
A possible formulation of the problem is as follows:
\begin{equation}\label{eq:quadrature-problem}
\begin{aligned}
  \bm{x}    &= [ \bm{w} ~ v_0 ]^T                             \quad &, \\
  R(\bm{x}) &= \| A(v_0) \bm{w} - \bm{b} \|                   \quad &, \\
  \min_{\bm{x} \in \Re} & ~ \frac{1}{2} R(\bm{x})^T R(\bm{x}) \quad &, \\
  s.t. ~ & R(\bm{x}) =  0                                     \quad &, \\
         & 0 < v_0 \leq v_{max}                               \quad &, \\
         & w_i \geq 0 ~ ~ \forall i                           \quad &.
\end{aligned}
\end{equation}
%
%
%



We have performed a detailed exploration of the available phase-space, implementing a solver 
for \autoref{eq:quadrature-problem} based on the \textit{LAPACK} library with several instances 
running in parallel on a cluster of CPUs. The solver takes as
input a stencil $G$ and tries to find a solution for \autoref{eq:quadrature-problem} 
by scanning several values of $v_0$ with a simple steepest-descent method.
This fast method allows to scan several stencils at different values of $m/T_R$; on the other hand, more 
robust techniques are desirable in order to perform a more systematic exploration of the phase-space.

\begin{figure}[t]
  \centering
  \includegraphics[width=.49\textwidth]{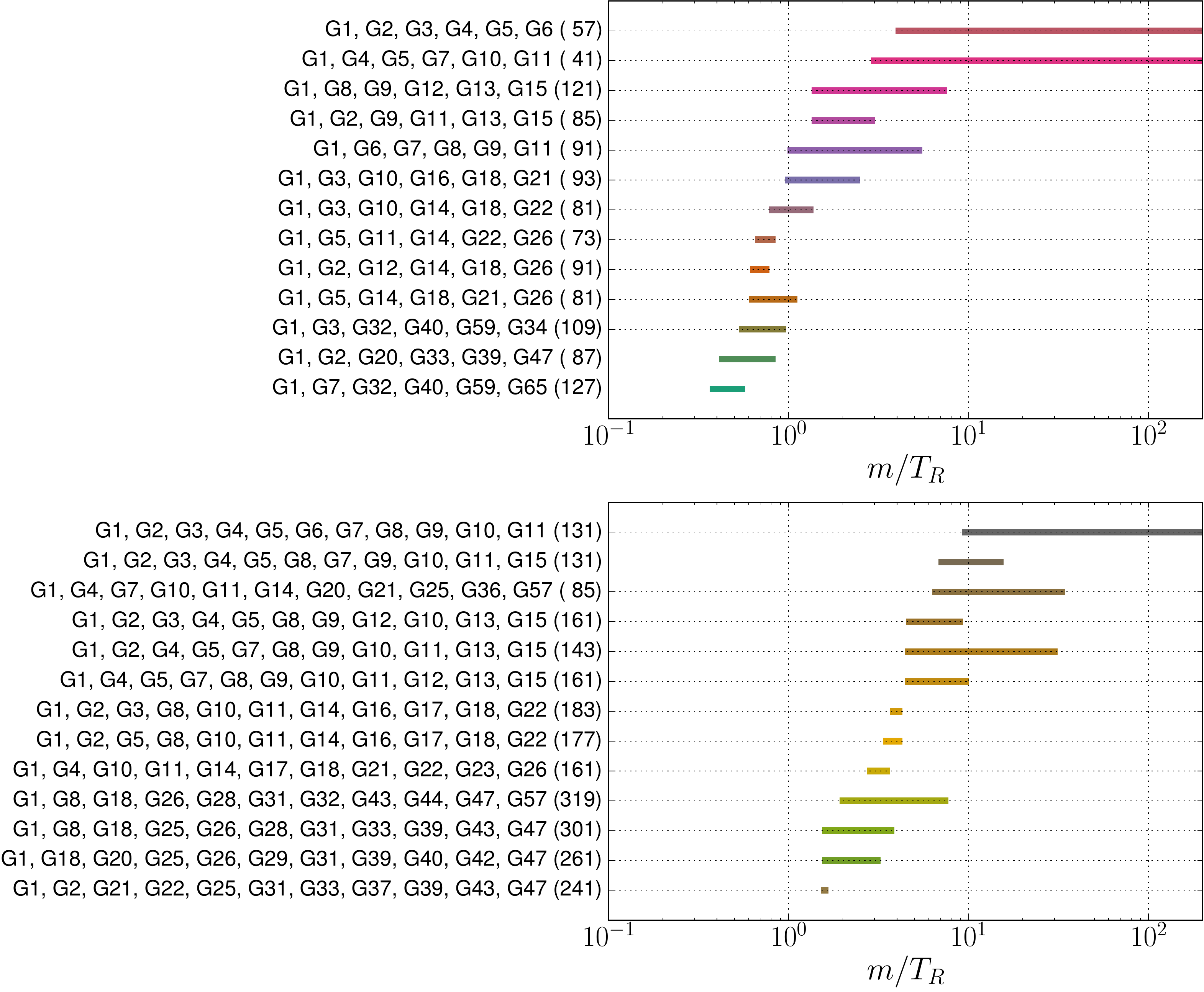}
  \caption{Stencils used to construct a numerically stable quadrature 
          for different value of $\bar{m}$. Top, stencils with six 
          velocity groups, for a second order approximation. 
          Bottom, stencils with eleven velocity vector groups, 
          for a third order approximation.
          Horizontal bars represent the working range of values $m / T_R$
          for each quadrature.}\label{fig:quadrature}
\end{figure}

Typically, for a given value of $\bar{m}$  several different stencils are possible; 
however, each stencil works correctly only in a certain range of $\bar{m}$.
Still, a reasonably small sets of stencils  allows to treat $\bar{m} \ge 0.35$ at the second 
order and $\bar{m} \ge 1.5$ at the third order, offering a possibility to cover a very large 
kinematic regime, from almost ultra-relativistic to non-relativistic. A graphical view of (a subset) 
of all stencils that we have identified, including the corresponding number of populations,  
is shown in \autoref{fig:quadrature} for both $2$-nd and $3$-rd order.

In general, the process of finding working quadratures becomes
harder  and harder as $\bar{m}$ takes smaller and smaller values.  The reason
for this, from a strictly mathematical point of view, is that for small  values
of $\bar{m}$ the condition number of the system matrix in
\autoref{eq:quadrature-problem}  takes large values, therefore requiring more
advanced linear algebra techniques. 
From a physical point of view the reason why this is a difficult problem, 
and in particular one cannot expect to find solutions for $\bar{m} = 0$, is that we require that 
different groups of particles travel in one time step at different distances hopping 
from a point of the grid to another point of the grid. 
In the close to ultra-relativistic regime this requires to restrict to stencils whose elements sit at the 
intersection between a Cartesian grid and a sphere of given radius. In this case, the trick used by   
Mendoza et al. \cite{mendoza-2013}, of using several energy 
shells, possible in the ultra-relativistic regime as velocity does not depend on energy, cannot be used if $\bar{m} \ne 0$.
Work is in progress to further clarify the best mathematical approach to finding the largest set of available solutions.

We have developed a plain \textit{C} program that implements our algorithm at second and third order.
The code is flexible enough to adapt to any of the possible quadratures and corresponding stencils.
The expressions for the polynomials and the equilibrium distribution, obtained using \textit{Mathematica} software, are
almost automatically translated into corresponding \textit{C} code lines. The overall structure of the code follows the
typical implementation of almost all LBM algorithms and shares the same opportunities for massive parallelization.
%
The number of floating point instructions per lattice site for one iteration of the collide kernel, 
at third order, is 83000 when employing a quadrature with 207 populations; 
this figure reduces down to 55000 using a 131 point stencil. Note that, consistently with the use of BGK-like 
collision operator, the instruction count scales approximately linearly with the number of discrete speeds 
and not quadratically as for the actual Boltzmann collision operator.
To put these numbers in perspective, the ultra-relativistic code in \cite{mendoza-2013} using a stencil with 128 
velocities requires about 52000 instructions; the slight difference can be accounted to the computation of 
the more complex equation of state and equilibrium function. 

\section{Numerical Validation}\label{sec:validation}

%
In this section we present a validation of the model by solving the Riemann problem for a quark-gluon plasma.
Such a choice is made in order to directly compare against previous RLBM formulations and other
relativistic hydrodynamics solvers dealing with the relativistic Boltzmann equation.
It should be noted that previous works have focused on the ultra-relativistic regime, which we can only
approximate using small values of $m/T_R$. As discussed in \autoref{sec:quadrature}
the minimum value of $m/T_R$ that can be used in simulations depends on whether we can find a stencil 
allowing to implement a quadrature for a given value of the rest mass.

In the following we compare different simulations for the 1D
shock-wave problem, showing that decreasing the value of $m / T$ our simulations
tends to the results of the ultra-relativistic regime, as computed by well-known
codes, such as ECHO-QGP \cite{rolando-2014}, the Boltzmann
approach multi-parton scattering (BAMPS) \cite{xu-greiner-2007}, and the
ultra-relativistic RLBM described in \cite{mendoza-2013}.

The initial conditions of the simulation, that follow a
benchmark performed by BAMPS, are defined by a pressure step having, in
physical units, $P_0 = 5.43 ~ GeV/fm^3$ and $P_1 = 0.339 ~ GeV/fm^3$, with
corresponding initial temperatures $T_0 = 400 ~ MeV$ and $T_1 = 200 ~
MeV$.

\begin{figure}[t]
  \includegraphics[width=\linewidth]{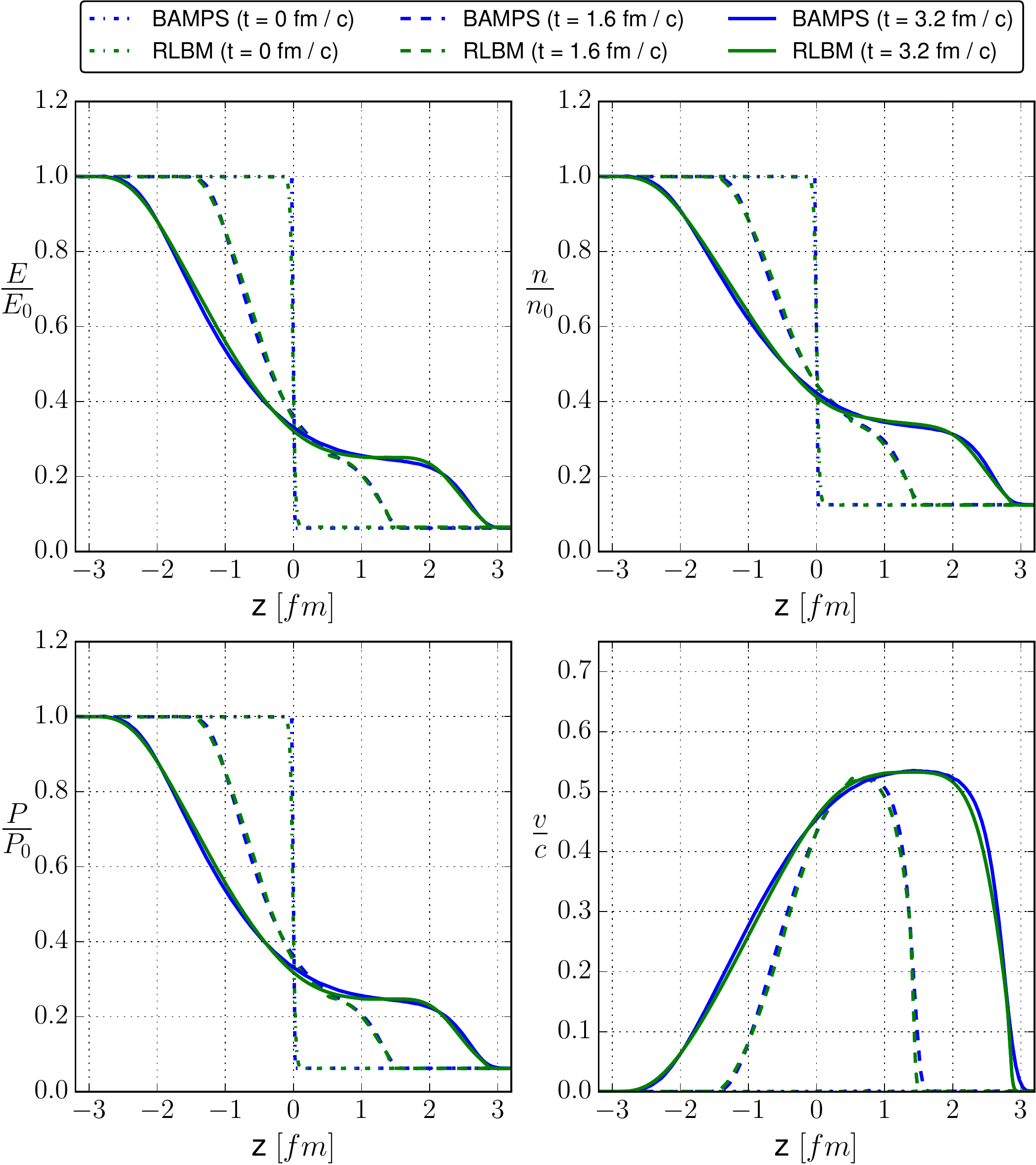}
  \caption{Comparison of the time-evolution of the solution of the Riemann
    problem obtained with BAMPS (blue lines) and with a second order RLBM solver
    using $m/T_R = 0.36$ (green lines).  We show the energy (top left),  density
    (top right), pressure (bottom left) and velocity of the shock wave (bottom
    right), at $t = 0~fm/c$, $t = 1.6~fm/c$ and $t = 3.2~fm/c$. We make use of a $133$ velocities
    stencil given by G = G1, G14, G49, G51, G60, G70.
    }\label{fig:bamps-rlbm}
\end{figure}
\begin{figure}[b]
  \includegraphics[width=\linewidth]{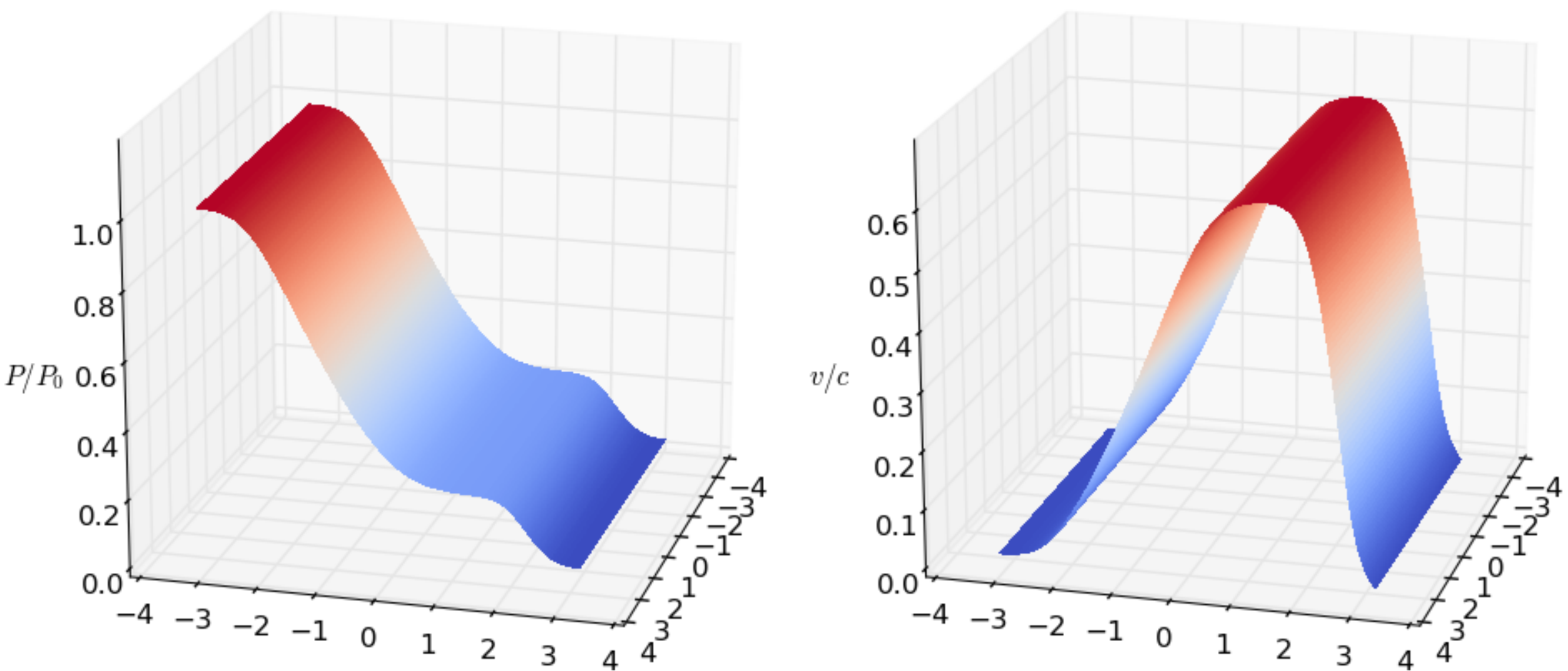}
  \caption{Pressure and velocity profiles (at $t = 3.2~fm/c$) of the same 1D Riemann problem of
  \autoref{fig:bamps-rlbm}, obtained via a 2D simulation. Results agree 
  with those of the 1D simulation to machine precision. }\label{fig:2dstep}
\end{figure}

To make contact with real-life physics,  it is necessary to convert from 
physical units to lattice units. In our simulations we set the
following values  for the initial temperature $T_0 = 1$, $T_1 = 0.5$ (this means
that we set the scale of our  reference temperature $T_R = 400 ~ MeV$), and
the values $n_0 = 1$, $n_1 = 0.124$  for the initial density, which correctly
reproduce the ratio $P_1 / P_0$.  
To relate physical space and time units with the corresponding lattice units, one starts by assigning the 
physical length $\delta x$ corresponding to one lattice spacing; one arbitrary population group in the simulation 
stencil, having velocity $||n_k|| v_0$, travels a distance of  $||n_k||$ lattice spacings in one time unit; 
then, if we call $dt$ the physical time unit corresponding to one discrete time step, 
$||n_k|| v_0 \delta t = ||n_k|| \delta x$, so we finally obtain $\delta t = \delta x / v_0$.

Another quantity that one must properly scale in order to use consistently different quadratures is
the relaxation time $\tau$. In the numerical setup $\tau$ is expressed in lattice time units, so it 
naturally follows from the discussion on the discrete time steps $\delta t$ that $\tau$ can be written as 
$\tau = \tau_{f} v_0 / \delta x$, and $\tau_{f}$ is related to the transport coefficients of the system that one wants to study.
\begin{figure}[t]
\includegraphics[width=\linewidth]{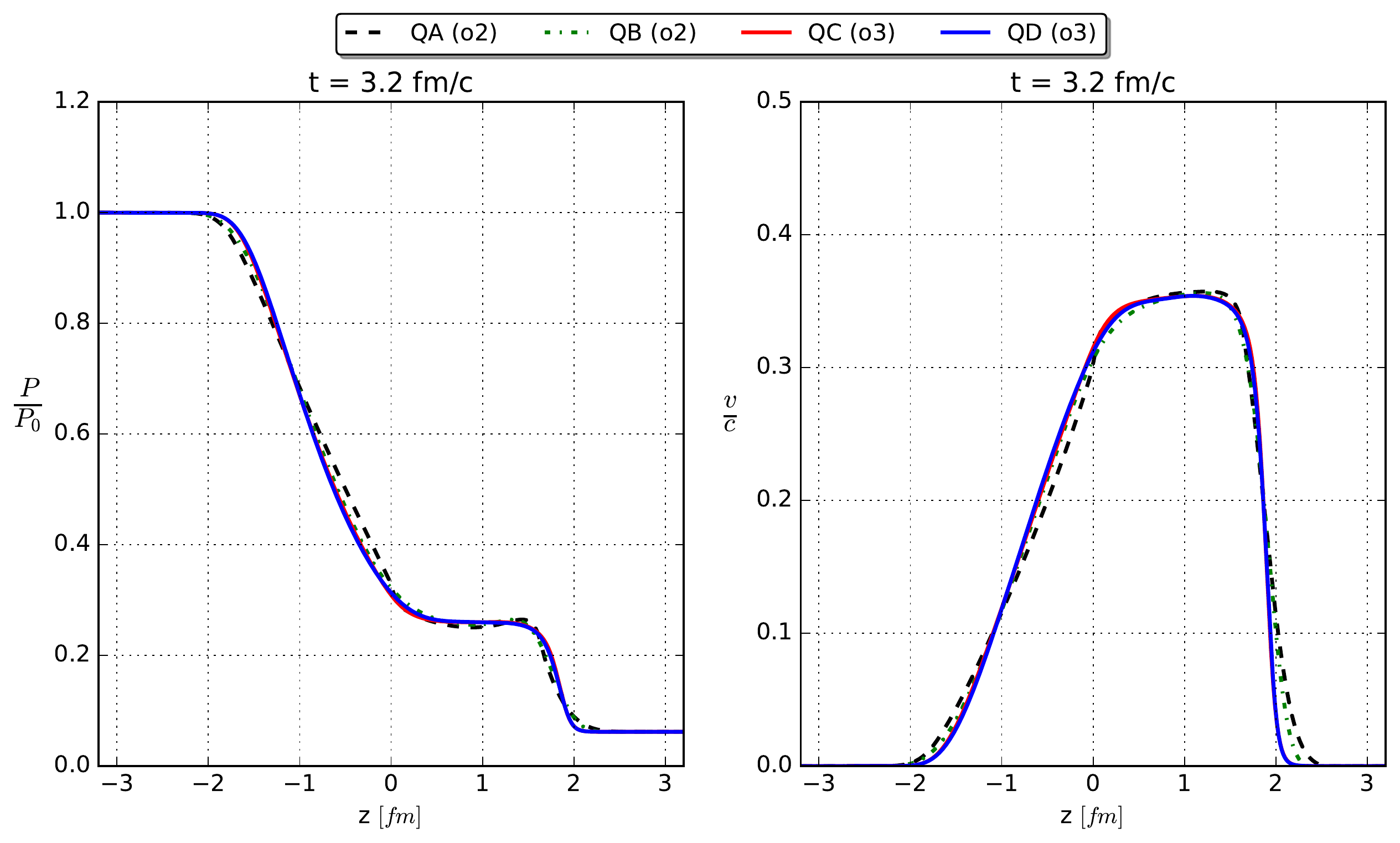}
\caption{Comparison of the time-evolution of the Riemann
         problem obtained with a RLBM solver using different quadratures, at the second and third order $(m/T_R = 5)$.
         Quadrature A ($2$-nd order): G = G1, G4, G5, G7, G10, G11.                            $v_0 = 0.2600738 $.
         Quadrature B ($2$-nd order): G = G1, G2, G3, G4,  G5,  G6.                            $v_0 = 0.3609900 $.
         Quadrature C ($3$-nd order): G = G1, G2,  G3,  G4,  G6,  G7,  G9, G10, G11, G13, G15. $v_0 = 0.2722674 $.
         Quadrature D ($3$-nd order): G = G1, G8, G18, G26, G28, G31, G32, G43, G44, G47, G57. $v_0 = 0.1571087 $. \\
         } \label{fig:cmp-timestep}
\end{figure}
The accurate link between the transport coefficients and $\tau_f$ in the relativistic regime is still 
debated in the literature. The approaches based on Grad's method of moments and on the Chapman-Enskog 
theory give slightly diverging results in the relativistic regime (even if they agree in the non-relativistic limit), 
see \cite{cercignani-2002}. Attempts to clarify this situation have been made by Israel and Stewart \cite{israel-1976} 
and more recently in a series of papers by Denicol et al. \cite{denicol-2012, molnar-2014}. In our tests, 
we use Grad's method (also adopted in \cite{mendoza-2013}), expecting only limited inaccuracies, of the order of $\simeq 10 - 15 \%$.
Therefore, at this stage we are not including contributions from massive particles in the transport coefficients,
thus relating the relaxation time $\tau$ to the shear viscosity $\eta$ through $ \eta = (2/3) P (\tau - \delta t / 2)$;
work is in progress in order to improve and generalize this definition for our RLBM formulation.
\begin{figure}[t]
\includegraphics[width=\linewidth]{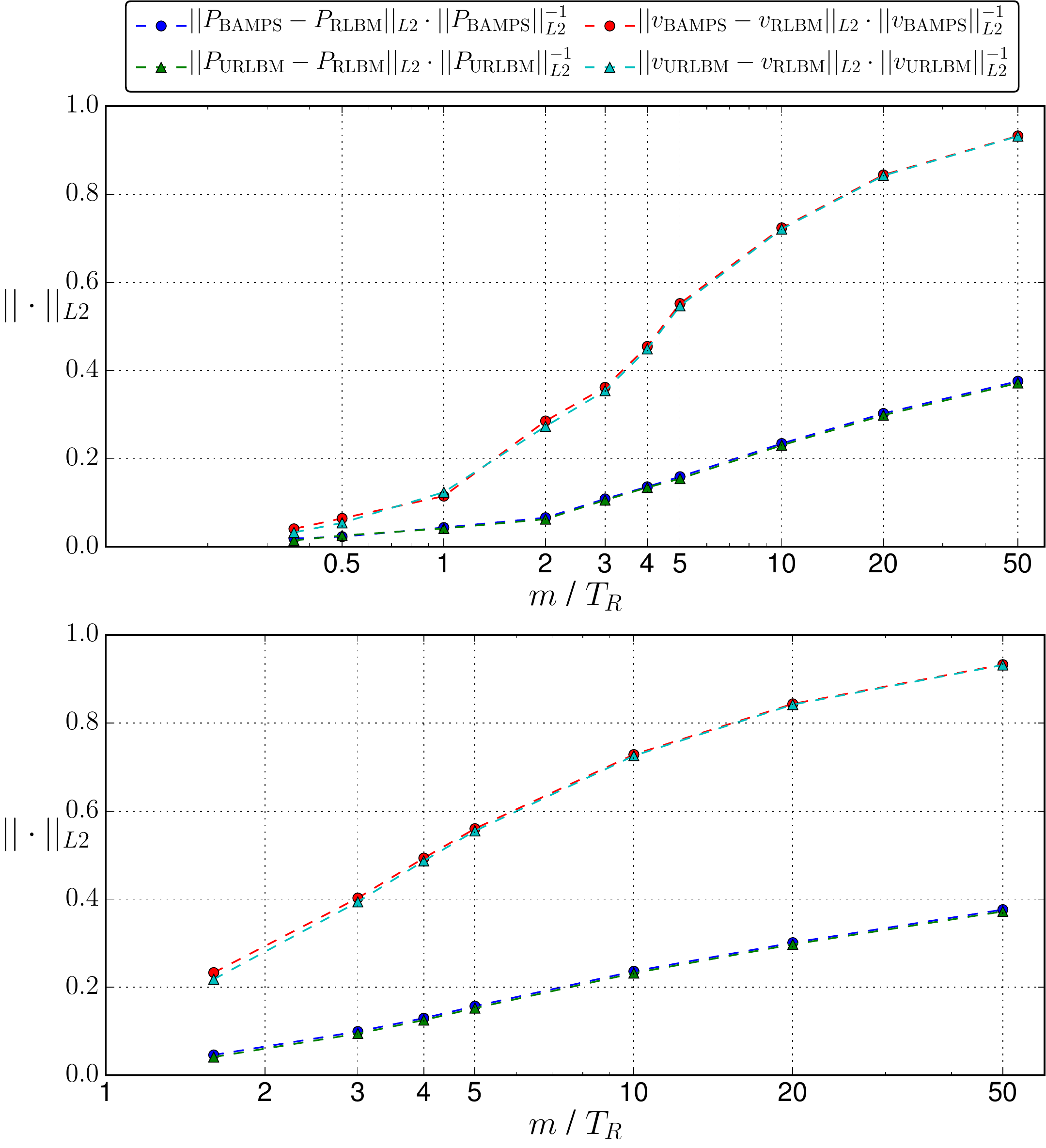}
\caption{Relative L2-distance of our simulated solution at second order (left) and third order (right)
with respect to BAMPS and with respect to the ultra-relativistic LBM (URLB), as a function of 
the $m/T_R$ ratio; we see that solutions become closer and closer as the $m/T_R$ ratio becomes smaller. 
Each point is obtained making use of the stencil having the smallest number of velocities.} \label{fig:L2err-o2o3}
\end{figure}
We perform our first tests on a lattice with $1\times1\times6400$ cells, half of which represents 
our domain defined in the interval $(-3.2~fm, 3.2~fm)$, and the other half forming a mirror which 
allows us to use periodic boundary conditions. It follows that on our grid $6.4 ~fm$ corresponds to  
$3200$ grid points, that is $\delta x = 0.002 ~fm$; the value of $\delta t $ for any given quadrature 
can be derived as explained in the previous paragraph.
\autoref{fig:bamps-rlbm} shows a typical result of this test, where we compare the energy, 
density, velocity and pressure profiles of BAMPS with $\eta / s = 0.1$ against our model.
Here $s$ is the entropy density, calculated according to the relation $ s = 4 n - n \ln{( n / n^{eq} )}$, where $n^{eq}$
comes from the equilibrium function, $n^{eq} = d_G T^3 / \pi^2$, with $d_G = 16 $ the degeneracy of the gluons.
The profiles show the evolution of the system from  $t = 0~fm/c$ to $t = 3.2~fm/c$; 
\autoref{fig:bamps-rlbm} shows that the results obtained simulating for $m/T_R = 0.36$ are 
in very good agreement with BAMPS; 
one may relate this nice and (possibly) unexpected behavior to the mild differences between the EOS 
of the two systems in this regime (see again 
\autoref{fig:edensity-ration}).

Although for validation purposes we  consider only one dimensional simulations, our
model readily extends to two and three spatial dimensions. \autoref{fig:2dstep} 
shows a 2D example where we solve the same Riemann problem on a 2D 
grid of $1\times800\times800$ sites; we have checked that 1D and 2D results agree almost to machine precision.

A further validation of our model is offered by the consistency among simulations of the same physical setup, 
performed with different quadratures.
In principle one expects the same results, after appropriately rescaling the space and time units as discussed 
above; in practice, small differences may appear, as different quadratures provide slightly different 
approximations to the distribution moments.

This is shown in \autoref{fig:cmp-timestep},
where we compare results obtained by simulating the same problem with different quadratures at the 2nd and 3rd order.
We see a close to perfect agreement at
third order. On the other hand, as one would expect, the results are slightly divergent when
using second order quadratures since the moments related to the viscosity terms are not fully recovered and different 
quadratures introduce different errors in their approximation.

We conclude this section with a more general validation test, presented in \autoref{fig:L2err-o2o3}, 
where we show that our solutions becomes closer and  closer to the ultra-relativistic
ones, as we reduce the $m/T$ ratio; this shows that the present algorithm is a 
good candidate to bridge the gap between ultra-relativistic and non-relativistic regimes.
%

\section{Results and prospective applications}\label{sec:results}

The RLBM scheme presented in this paper allows to explore many different physics regimes, from a 
nearly ultra-relativistic behavior, down to mildly relativistic ones. 
While we leave physics applications to future works, here we wish to  offer  just a few preliminary
examples,  conveying a sense of prospective physics applications of the present method.

\begin{figure}[t]
\centering
\includegraphics[width=\linewidth]{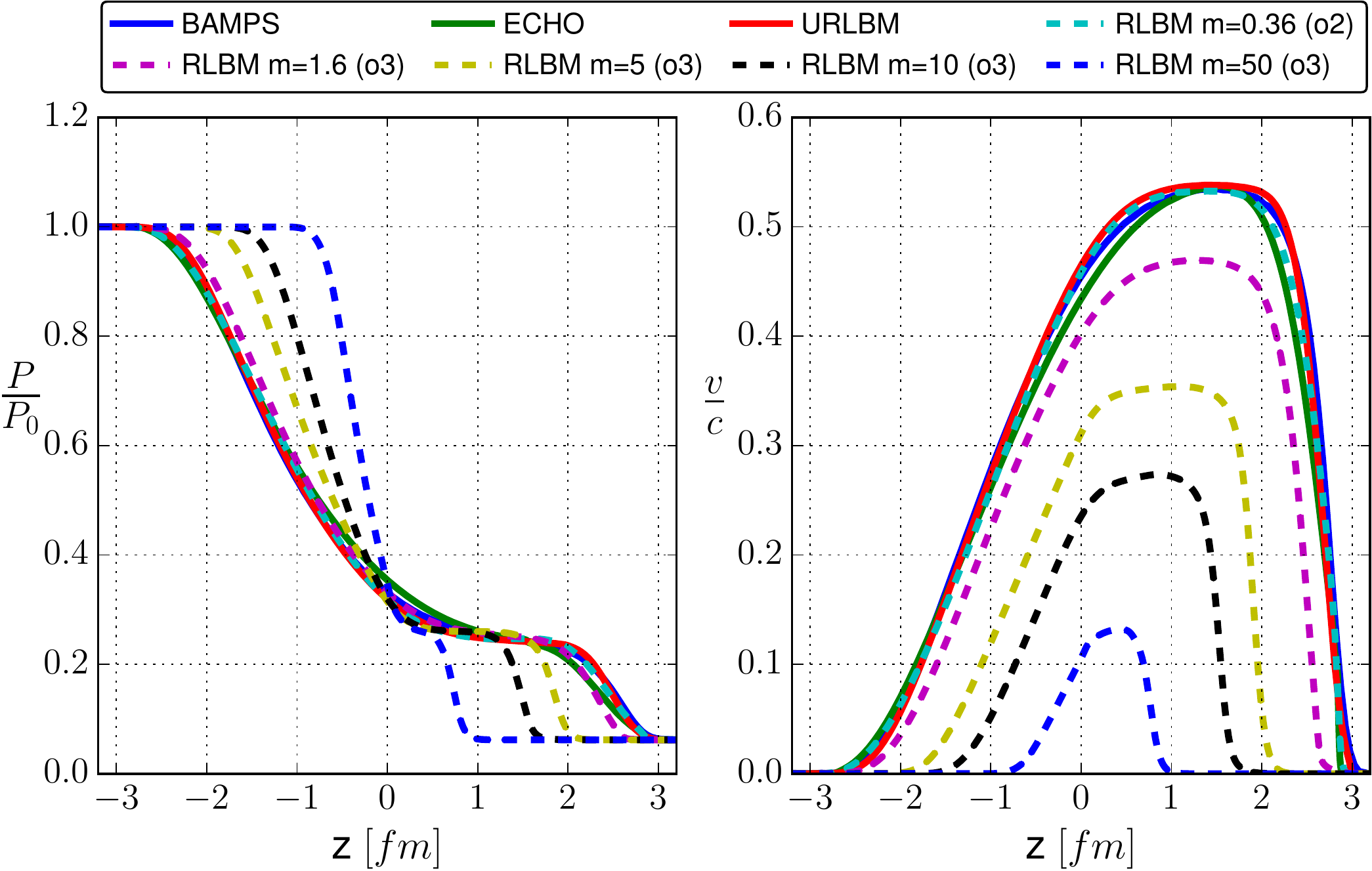}
\caption{Comparison of different relativistic hydrodynamics solvers.
   Dashed lines represent the results obtained using our RLBM solver 
   implementing a second and a third order approximation for different values of $\bar{m}$.
Left) Pressure profile at $t = 3.2 ~fm/c$.
Right) Velocity profile at $t = 3.2 ~fm/c$.
}\label{fig:pressure-velocity-cmp}
\end{figure}

In \autoref{fig:pressure-velocity-cmp} we show the behavior of a fluid undergoing a Riemann-like shock,
for a fixed value of $\eta / s$, and for different values of $m / T$ placing ourselves at different relativistic regimes.
One easily appreciates the changes in the system evolution as one moves from a 
strongly relativistic to an almost classic regime: the evolution from the same initial temperature and pressure gradients becomes slower as $m / T$ becomes larger.

We conclude by testing the stability of our algorithm when considering fully 
two-dimensional simulations. Once again, the reader should be aware that the
aim  here is not a detailed study of a real physical application but rather to
give  a taste of what the model allows to do. We also like to point out that
while the  examples presented in the following have been performed also in three
dimensions,  we report here only two dimensional profiles for the sake of
visualization.

\begin{figure}[t]
\begin{subfigure}{0.49\textwidth}
\includegraphics[width=\linewidth]{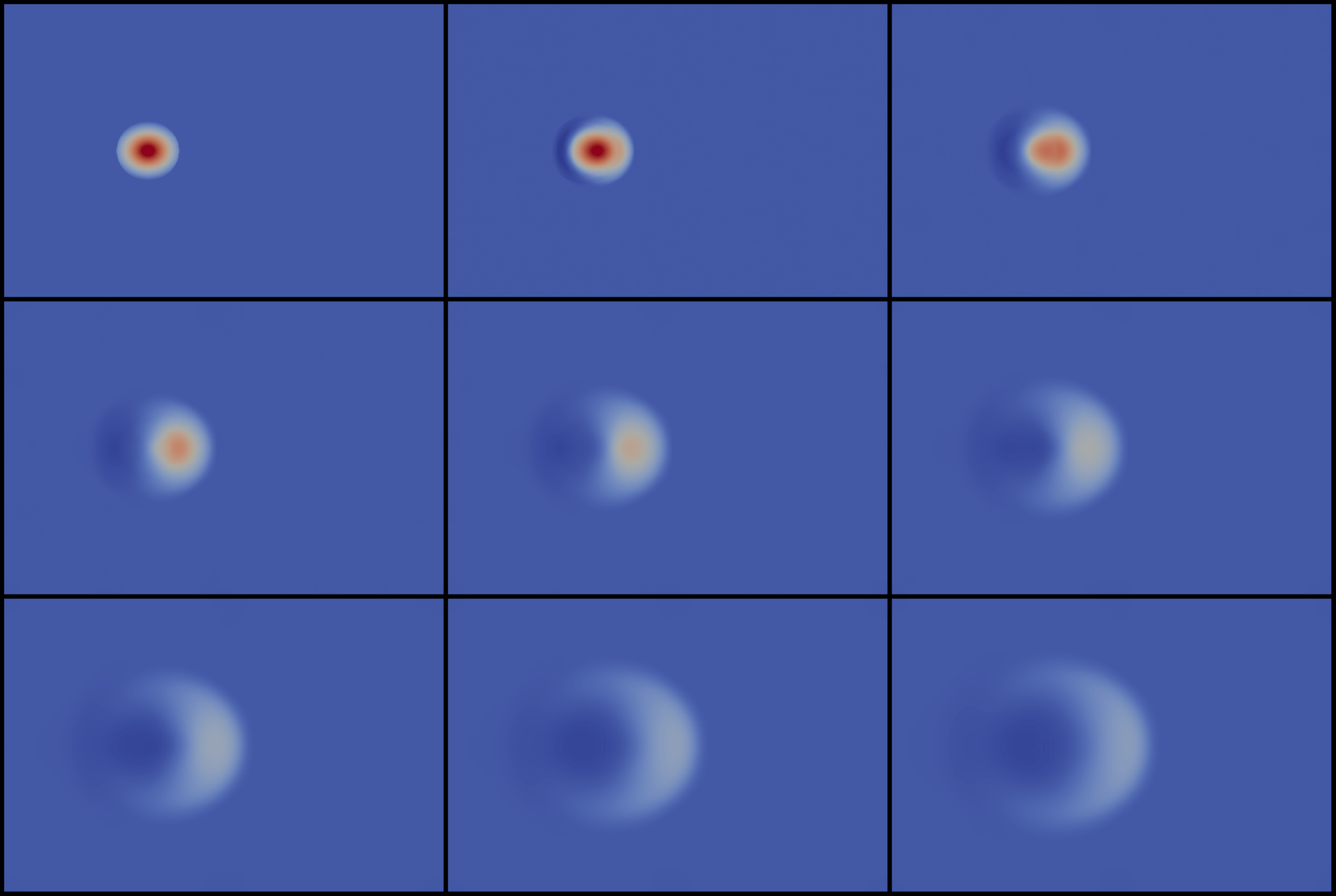}
\caption{Density} \label{fig:density-ex1}
\end{subfigure}
\hspace*{\fill} 
\begin{subfigure}{0.49\textwidth}
\includegraphics[width=\linewidth]{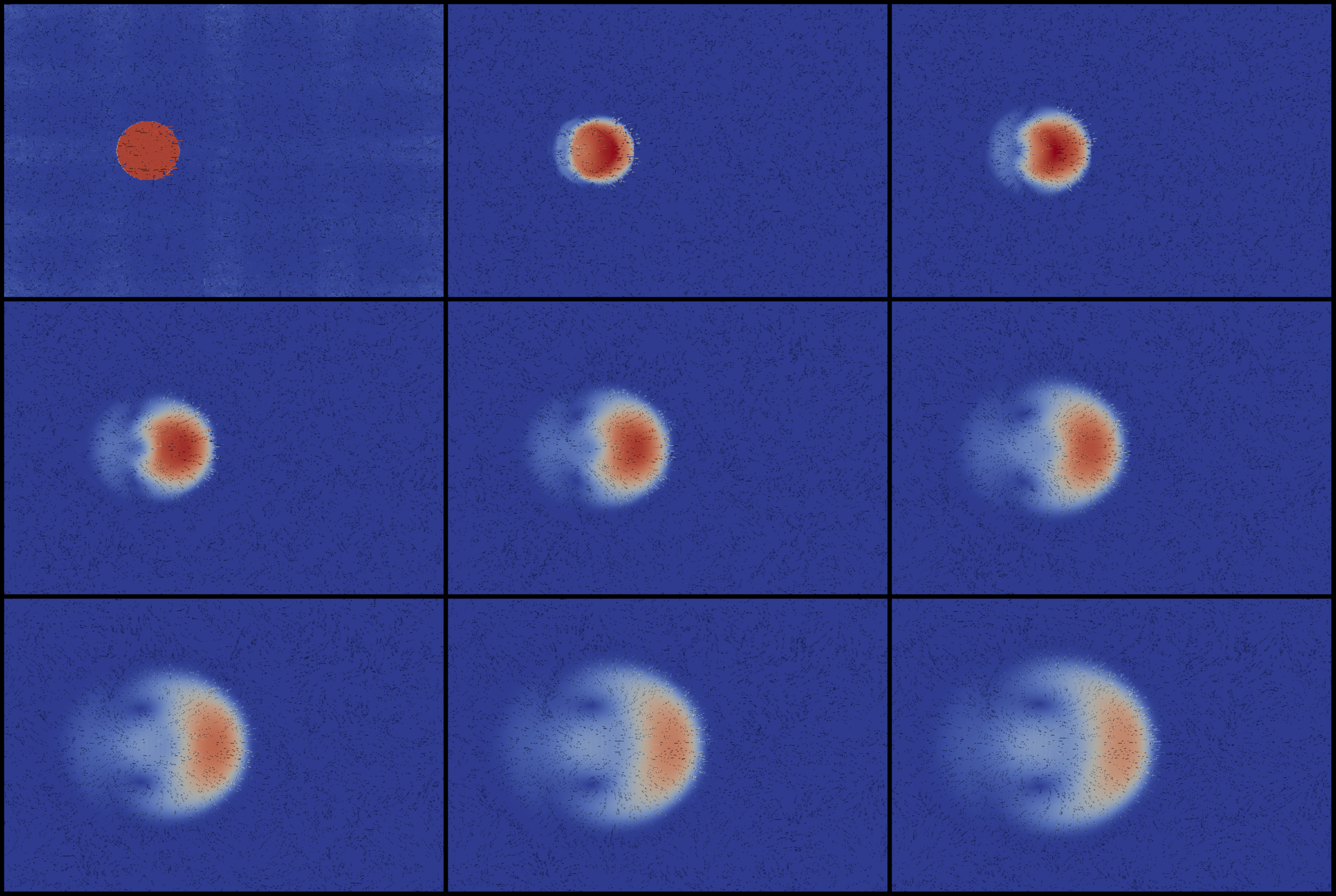}
\caption{Velocity} \label{fig:velocity-ex1}
\end{subfigure}
\caption{ Evolution of a Lorentz contracted circular domains with a radial initial density moving with a initial velocity $v = 0.5 ~ c$.
          Starting from $t=0$ we present 9 frames, taken at $6$ time steps apart, showing profiles of: 
          a) Particle density. 
          b) Velocity magnitude.
        }\label{fig:2d-ex1}
\end{figure}

In \autoref{fig:2d-ex1} we present a relativistic analogue of  the Taylor-dispersion process \cite{taylor-1953}, simulating 
the dynamics of a circular domain of radius $R$ with 
an initial radial density $n(r) = n_0 ( 1 - r/R )$, and an initial velocity 
$v = (\frac{c}{2}, 0)$. \autoref{fig:density-ex1} outlines the density profile of the system at eight different 
time steps, while in parallel \autoref{fig:velocity-ex1}
shows the evolution of the velocity magnitude; the pictures show a substantial mixing of the 
fast moving fluid with the environment, qualitatively similar to the one exhibited by its low-speed analogue.

\begin{figure}[H]
\begin{subfigure}{0.49\textwidth}
\includegraphics[width=\linewidth]{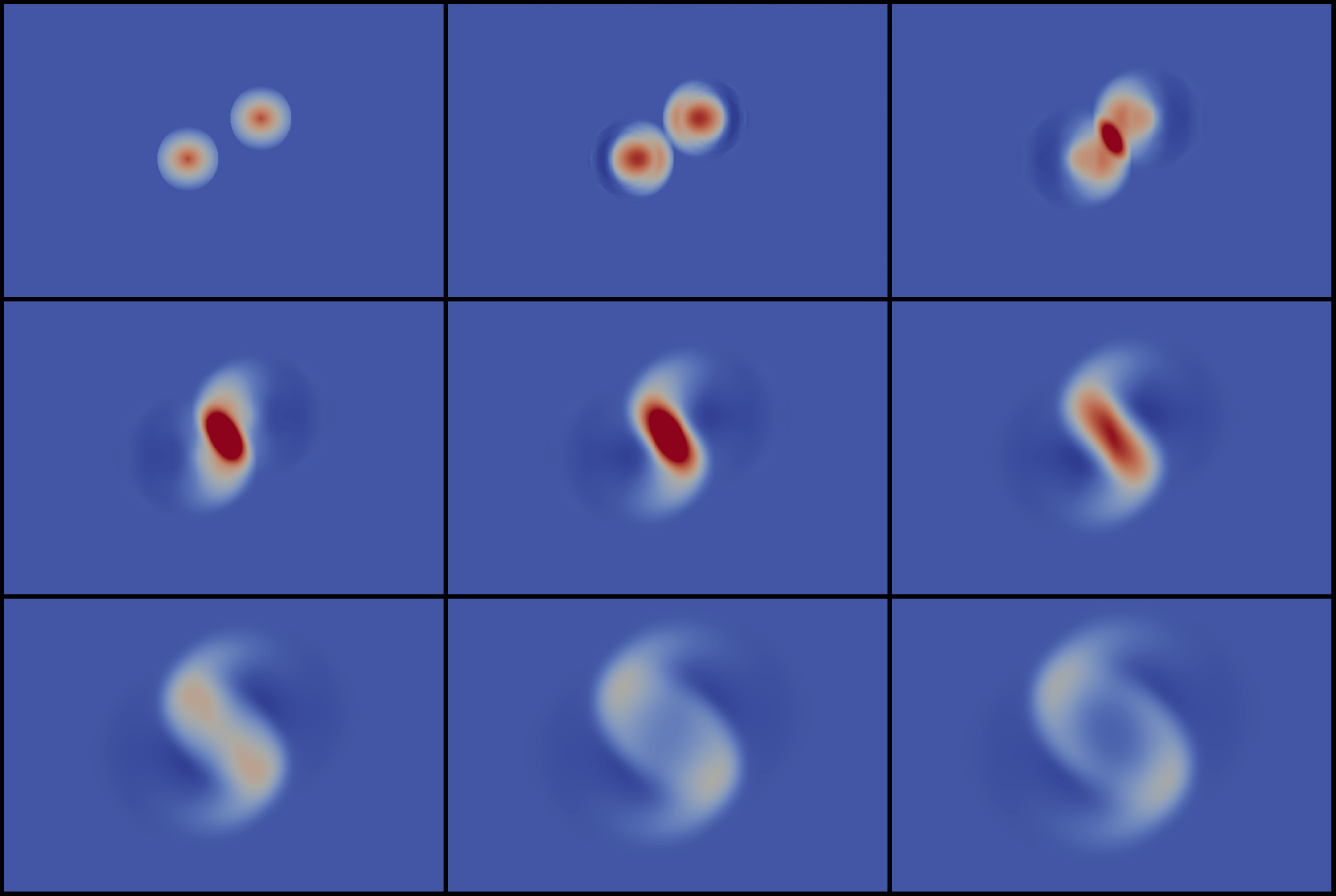}
\caption{Pressure} \label{fig:pressure-ex2}
\end{subfigure}
\hspace*{\fill} 
\begin{subfigure}{0.49\textwidth}
\includegraphics[width=\linewidth]{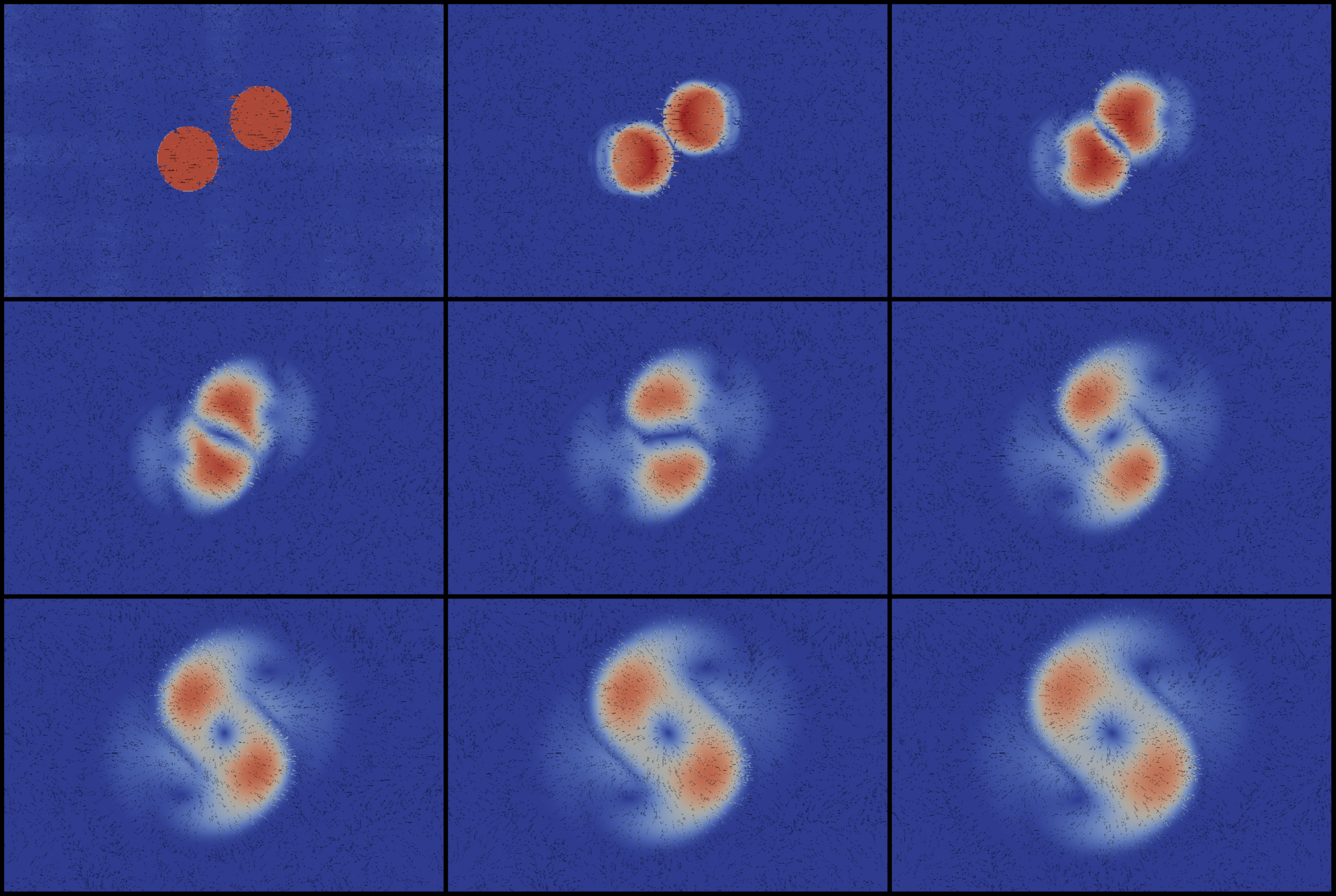}
\caption{Velocity} \label{fig:velocity-ex2}
\end{subfigure}
\caption{ Evolution of the collision between two Lorentz contracted circular domains with 
          a radial initial density moving with a initial velocity $v = 0.5 ~ c$.
          Starting from $t=0$ we present 9 frames, taken at $6$ time steps apart, showing profiles of: 
          a) Pressure. 
          b) Velocity magnitude.
        }\label{fig:2d-ex2}
\end{figure}

The simulation presented in \autoref{fig:2d-ex2} points in the direction of quark-gluon plasma phenomenology, 
as we consider two domains with the same initial density as before, traveling in opposite directions at
speed $v \sim \frac{c}{2}$. The pressure profile (\autoref{fig:pressure-ex2}),
and the velocity magnitude (\autoref{fig:velocity-ex2}) are shown at eight different time steps. 
%
%

In order to apply our model to realistic simulations in quark-gluon plasma
one would need to consider further extensions, as for instance suitable
EOS and appropriate initial and boundary conditions.
Still, the picture capture features of an hydrodynamic evolution of fluid
following the initial collision among nuclei,  as long as the temperature
remains larger than the freeze-out threshold \cite{cooper-1974}. Our
simulation method might also permit to apply the fluid-dynamic approach
across the deconfinement transition point, in the very interesting region
where the fluid speed of ``sound'' ($c^2_s = \partial p / \partial \epsilon$) 
changes abruptly (decreasing down to the so-called ``softest point'' 
and then increasing again)  and in which the role of the hadronic
degrees of freedom will start to play a significant dynamic role \cite{heinz-2005},\cite{chojnacki-2005}.
We expect the range $1 < m/T < 5$ to be relevant in
this case, as they correspond to the ratio of the mass of the low-lying
hadrons (e.g. pion, $\rho$-meson, nucleon) to the deconfinement temperature.
%
%
%
\section{Conclusions and Outlook}\label{sec:conclusions}
In this paper we have developed a new
class of relativistic lattice Boltzmann methods, that provides a wider
degree of generality and flexibility with respect to previous models, while still 
preserving all features of conceptual simplicity and computational
efficiency of lattice Boltzmann methods.
The main results of this work are summarized as follows:
\begin{itemize}
\item We have explicitly built a new class of RLBM based on massive pseudo-particles, 
able to recover the moments of the relativistic equilibrium distribution up to third order.
\item The use of massive pseudo-particles translates into the possibility
to tailor the detailed features of the method to fit a specific
relativistic range of velocities of the simulated system, ranging from
strongly relativistic to almost classical.
\item We have established a methodology capable of quickly deriving many
different variants of the present RLBM, allowing for instance to use
different sets of pseudo-particles and to adjust the value of $m/T$.
\item The algorithmic structure of the present RLBM is very similar to that
of other established lattice Boltzmann methods and the computational
complexity is not much higher. 
%
This algorithm retains the same computational advantages, offering high amenability to
parallelization, that can be exploited to write efficient high-performance computing codes.
\item Initial tests have shown that our algorithms are computationally
stable and robust over a wide range of physical parameters.
\end{itemize}
To the best of our knowledge, this is the first RLBM implementing
exact streaming on a Cartesian lattice without losing spatial resolution
and still recovering higher order moments of the equilibrium distribution.
The flexibility of this RLBM should make it an appealing computational tool 
to study several relevant relativistic physics problems, including for instance astrophysical 
contexts \cite{shore-1992} or quark-gluon plasma dynamics \cite{ackermann-2001}, or the transport properties 
of electronic pseudo-particles in 2D or 3D solid-state systems \cite{muller-2008, wan-2011}.
From the algorithmic point of view, we plan to work on further optimizations of the method, by studying for instance the physics 
accuracy and the computational efficiency of different stencil options.
We also plan to look at the low-velocity limit of our algorithms, with the goal of fully bridging the gap between 
relativistic and non-relativistic LBMs. 
Work along these lines is in progress.

\section*{Acknowledgement}
We thank M. Sbragaglia and P. Mocz for useful discussions.
AG has been supported by the European Union's Horizon 2020 research and
innovation programme under the Marie Sklodowska-Curie grant agreement No. 642069.
MM and SS thank the European Research Council (ERC) Advanced Grant No. 319968-FlowCCS
for financial support. 

\onecolumngrid
\appendix
\renewcommand{\thesection}{A}
\section{Second-order Relativistic Orthonormal Polynomials}\label{sec:appendixA}
This and the following appendix sections provide several mathematical details relevant in our calculations. As mathematical complexity quickly becomes very large, many very complex expressions are available as supplemental information \cite{supplementary-material}.
In this section we provide the analytic expressions of the orthonormal polynomials, up to
the second order, as a function of $\bar{p}^{\mu} =  p^{\mu} / T_R$.
\begin{align*}
J^{(0)}      &= 1 \\
J^{(1)}_0    &= \frac{1}{A} \left(\bar{p}_{0} -\frac{\bar{m} K_2(\bar{m})}{K_1(\bar{m})} \right) \\
J^{(1)}_x    &= \frac{1}{B} \bar{p}_{x}\\
J^{(1)}_y    &= \frac{1}{B} \bar{p}_{y}\\
J^{(1)}_z    &= \frac{1}{B} \bar{p}_{z}\\
J^{(2)}_{00} &= \frac{1}{\sqrt{3} \ C} \left( \bar{p}_{0}^2 + \bar{p}_{0} \left(\frac{3}{-\frac{\bar{m} K_1(\bar{m})}{K_2(\bar{m})}+\frac{\bar{m} K_2(\bar{m})}{K_1(\bar{m})}-3}-3\right)+\frac{3 K_1(\bar{m}) (\bar{m} K_1(\bar{m})+3 K_2(\bar{m}))}{\bar{m} K_1(\bar{m}){}^2+3 K_2(\bar{m}) K_1(\bar{m})-\bar{m} K_2(\bar{m}){}^2}-\bar{m}^2-3 \right) \\
J^{(2)}_{0x} &= \frac{1}{D} \bar{p}_{x} \left( \bar{p}_{0} -\frac{\bar{m} K_3(\bar{m})}{K_2(\bar{m}) }  \right) \\
J^{(2)}_{0y} &= \frac{1}{D} \bar{p}_{y} \left( \bar{p}_{0} -\frac{\bar{m} K_3(\bar{m})}{K_2(\bar{m}) }  \right) \\
J^{(2)}_{0z} &= \frac{1}{D} \bar{p}_{z} \left( \bar{p}_{0} -\frac{\bar{m} K_3(\bar{m})}{K_2(\bar{m}) }  \right) \\
J^{(2)}_{xx} &= \frac{1}{E} \left( \frac{\bar{p}_{y}^2}{2 } -\frac{\bar{p}_{z}^2}{2 }\right)\\
J^{(2)}_{yy} &= \frac{1}{2 \sqrt{3} E} \left( -2 \bar{p}_{x}^2 + \bar{p}_{y}^2 + \bar{p}_{z}^2 \right) \\
J^{(2)}_{xz} &= \frac{1}{E} \bar{p}_{x} \bar{p}_{z}\\
J^{(2)}_{yz} &= \frac{1}{E} \bar{p}_{y} \bar{p}_{z}\\
J^{(2)}_{xy} &= \frac{1}{E} \bar{p}_{x} \bar{p}_{y}\\
\end{align*}
with:
\begin{align*}
A &= \sqrt{\bar{m} \left(\bar{m}+\frac{K_2(\bar{m}) (3 K_1(\bar{m})-\bar{m} K_2(\bar{m}))}{K_1(\bar{m}){}^2}\right)} \\
B &= \sqrt{\frac{\bar{m} K_2(\bar{m})}{K_1(\bar{m})}} \\
C &= \sqrt{2 \bar{m}^2+\frac{5 \bar{m} K_2(\bar{m})}{K_1(\bar{m})}-\frac{3 K_1(\bar{m}) (\bar{m} K_1(\bar{m})+3 K_2(\bar{m}))}{\bar{m} K_1(\bar{m}){}^2+3 K_2(\bar{m}) K_1(\bar{m})-\bar{m} K_2(\bar{m}){}^2}+3}  \\
D &= \sqrt{\frac{\bar{m}^2 \left(\bar{m} K_2(\bar{m}){}^2+5 K_3(\bar{m}) K_2(\bar{m})-\bar{m} K_3(\bar{m}){}^2\right)}{K_1(\bar{m}) K_2(\bar{m})}} \\
E &= \sqrt{\frac{\bar{m}^2 K_3(\bar{m})}{K_1(\bar{m})}} \\
\end{align*}

\newpage
\appendix
\renewcommand{\thesection}{B}
\section{Second-order Orthogonal Projections}\label{sec:appendixB}
In this appendix we will provide the analytic expressions of the 
orthogonal projections $a^{(k)}$, up to the second order, written as
$$a^{(k)} = \frac{1}{T_R} b^{(k)} $$

\begin{align*}
b^{(0)}      &= \frac{1}{\bar{m}} \frac{K_1( \frac{m}{T} ) }{  K_2( \frac{m}{T} ) }  \\ 
b^{(1)}_0    &= \frac{1}{A} \left( U_0 - \frac{K_2(\bar{m})}{K_1(\bar{m})} \frac{ K_1( \frac{m}{T} ) }{ K_2( \frac{m}{T} )} \right)\\
b^{(1)}_x    &= \frac{1}{B} U_x \\
b^{(1)}_y    &= \frac{1}{B} U_y \\
b^{(1)}_z    &= \frac{1}{B} U_z \\
b^{(2)}_{00} &= \frac{1}{\sqrt{3} C} \left( U_0^2 \left( \bar{m} \frac{K_3( \frac{m}{T} )}{K_2( \frac{m}{T} )}  \right) + U_0 \left( \frac{1}{-\frac{\bar{m} K_1(\bar{m})}{K_2(\bar{m})}+\frac{\bar{m} K_2(\bar{m})}{K_1(\bar{m})}-3}-1 \right) \right. \\
            & \left.    + \frac{ K_1( \frac{m}{T} ) }{ K_2( \frac{m}{T} )} \left( \frac{-\bar{m}^2 K_1(\bar{m})^2+\left(\bar{m}^2+3\right) K_2(\bar{m})^2-3 \bar{m} K_2(\bar{m}) K_1(\bar{m})}{\bar{m} K_1(\bar{m})^2+3 K_2(\bar{m}) K_1(\bar{m})-\bar{m} K_2(\bar{m})^2} + \frac{T}{T_R} \frac{K_2( \frac{m}{T} )}{K_1( \frac{m}{T} )}  \right) \right) \\
b^{(2)}_{0x} &= \frac{1}{D} \left( \bar{m} \frac{K_3( \frac{m}{T} )}{K_2( \frac{m}{T} )} U_0 U_x - \bar{m} \frac{ K_3( \bar{m} ) }{ K_2( \bar{m} ) } U_x  \right) \\
b^{(2)}_{0y} &= \frac{1}{D} \left( \bar{m} \frac{K_3( \frac{m}{T} )}{K_2( \frac{m}{T} )} U_0 U_y - \bar{m} \frac{ K_3( \bar{m} ) }{ K_2( \bar{m} ) } U_y  \right) \\
b^{(2)}_{0z} &= \frac{1}{D} \left( \bar{m} \frac{K_3( \frac{m}{T} )}{K_2( \frac{m}{T} )} U_0 U_z - \bar{m} \frac{ K_3( \bar{m} ) }{ K_2( \bar{m} ) } U_z  \right) \\
b^{(2)}_{xx} &= \frac{1}{2 \sqrt{3} E} \bar{m} \frac{ K_3( \frac{m}{T} ) }{ K_2( \frac{m}{T}) }  \left( -  2 U_x^2 + U_y^2 + U_z^2  \right)\\
b^{(2)}_{yy} &= \frac{1}{2 E} \bar{m} \frac{ K_3( \frac{m}{T} ) }{ K_2( \frac{m}{T} ) } \left(    U_y^2  -  U_z^2 \right)\\
b^{(2)}_{xz} &= \frac{1}{E} \left( \bar{m} \frac{ K_3( \frac{m}{T} ) }{ K_2( \frac{m}{T} ) } U_x U_z \right)\\
b^{(2)}_{yz} &= \frac{1}{E} \left( \bar{m} \frac{ K_3( \frac{m}{T} ) }{ K_2( \frac{m}{T} ) } U_x U_y \right)\\
b^{(2)}_{xy} &= \frac{1}{E} \left( \bar{m} \frac{ K_3( \frac{m}{T} ) }{ K_2( \frac{m}{T} ) } U_y U_z \right)
\end{align*}

\appendix
\renewcommand{\thesection}{C}
\section{Stencils for three-dimensional lattices}\label{sec:appendixC}

\begin{center}
\begin{table}[H] 
\centering 
\resizebox{0.95\textwidth }{!}{
  \begin{tabular}[t]{|c|c|c|rllr|}
  \hline
  Group & Vectors & \# Vectors & \multicolumn{4}{c|}{Length}\\
  \hline
  \hline
   1 &  $(\phantom{\pm} 0, \phantom{\pm} 0, \phantom{\pm} 0)_{\texttt{FS}}$ &   1 & $          0$ & ( = & 0.      & ) \\
   2 &  $(         \pm  1, \phantom{\pm} 0, \phantom{\pm} 0)_{\texttt{FS}}$ &   6 & $          1$ & ( = & 1.      & ) \\
   3 &  $(         \pm  1,          \pm  1, \phantom{\pm} 0)_{\texttt{FS}}$ &  12 & $  \sqrt{ 2}$ & ( = & 1.41421 & ) \\
   4 &  $(         \pm  1,          \pm  1,          \pm  1)_{\texttt{FS}}$ &   8 & $  \sqrt{ 3}$ & ( = & 1.73205 & ) \\
   5 &  $(         \pm  2, \phantom{\pm} 0, \phantom{\pm} 0)_{\texttt{FS}}$ &   6 & $          2$ & ( = & 2.      & ) \\
   6 &  $(         \pm  2,          \pm  1, \phantom{\pm} 0)_{\texttt{FS}}$ &  24 & $  \sqrt{ 5}$ & ( = & 2.23607 & ) \\
   7 &  $(         \pm  2,          \pm  2, \phantom{\pm} 0)_{\texttt{FS}}$ &  12 & $2 \sqrt{ 2}$ & ( = & 2.82843 & ) \\
   8 &  $(         \pm  2,          \pm  1,          \pm  1)_{\texttt{FS}}$ &  24 & $  \sqrt{ 6}$ & ( = & 2.44949 & ) \\
   9 &  $(         \pm  2,          \pm  2,          \pm  1)_{\texttt{FS}}$ &  24 & $          3$ & ( = & 3.      & ) \\
  10 &  $(         \pm  2,          \pm  2,          \pm  2)_{\texttt{FS}}$ &   8 & $2 \sqrt{ 3}$ & ( = & 3.4641  & ) \\
  11 &  $(         \pm  3, \phantom{\pm} 0, \phantom{\pm} 0)_{\texttt{FS}}$ &   6 & $          3$ & ( = & 3.      & ) \\
  12 &  $(         \pm  3,          \pm  1, \phantom{\pm} 0)_{\texttt{FS}}$ &  24 & $  \sqrt{10}$ & ( = & 3.16228 & ) \\
  13 &  $(         \pm  3,          \pm  2, \phantom{\pm} 0)_{\texttt{FS}}$ &  24 & $  \sqrt{13}$ & ( = & 3.60555 & ) \\
  14 &  $(         \pm  3,          \pm  3, \phantom{\pm} 0)_{\texttt{FS}}$ &  12 & $3 \sqrt{ 2}$ & ( = & 4.24264 & ) \\
  15 &  $(         \pm  3,          \pm  1,          \pm  1)_{\texttt{FS}}$ &  24 & $  \sqrt{11}$ & ( = & 3.31662 & ) \\
  16 &  $(         \pm  3,          \pm  2,          \pm  1)_{\texttt{FS}}$ &  48 & $  \sqrt{14}$ & ( = & 3.74166 & ) \\
  17 &  $(         \pm  3,          \pm  2,          \pm  2)_{\texttt{FS}}$ &  24 & $  \sqrt{19}$ & ( = & 4.3589  & ) \\
  18 &  $(         \pm  3,          \pm  3,          \pm  1)_{\texttt{FS}}$ &  24 & $  \sqrt{17}$ & ( = & 4.12311 & ) \\
  19 &  $(         \pm  3,          \pm  3,          \pm  2)_{\texttt{FS}}$ &  24 & $  \sqrt{22}$ & ( = & 4.69042 & ) \\
  20 &  $(         \pm  3,          \pm  3,          \pm  3)_{\texttt{FS}}$ &   8 & $3 \sqrt{ 3}$ & ( = & 5.19615 & ) \\
  21 &  $(         \pm  4, \phantom{\pm} 0, \phantom{\pm} 0)_{\texttt{FS}}$ &   6 & $          4$ & ( = & 4.      & ) \\
  22 &  $(         \pm  4,          \pm  1, \phantom{\pm} 0)_{\texttt{FS}}$ &  24 & $  \sqrt{17}$ & ( = & 4.12311 & ) \\
  23 &  $(         \pm  4,          \pm  2, \phantom{\pm} 0)_{\texttt{FS}}$ &  24 & $2 \sqrt{ 5}$ & ( = & 4.47214 & ) \\
  24 &  $(         \pm  4,          \pm  3, \phantom{\pm} 0)_{\texttt{FS}}$ &  24 & $          5$ & ( = & 5.      & ) \\
  25 &  $(         \pm  4,          \pm  4, \phantom{\pm} 0)_{\texttt{FS}}$ &  12 & $4 \sqrt{ 2}$ & ( = & 5.65685 & ) \\
  26 &  $(         \pm  4,          \pm  1,          \pm  1)_{\texttt{FS}}$ &  24 & $3 \sqrt{ 2}$ & ( = & 4.24264 & ) \\
  27 &  $(         \pm  4,          \pm  2,          \pm  1)_{\texttt{FS}}$ &  48 & $  \sqrt{21}$ & ( = & 4.58258 & ) \\
  28 &  $(         \pm  4,          \pm  3,          \pm  1)_{\texttt{FS}}$ &  48 & $  \sqrt{26}$ & ( = & 5.09902 & ) \\
  29 &  $(         \pm  4,          \pm  4,          \pm  1)_{\texttt{FS}}$ &  24 & $  \sqrt{33}$ & ( = & 5.74456 & ) \\
  30 &  $(         \pm  4,          \pm  2,          \pm  2)_{\texttt{FS}}$ &  24 & $2 \sqrt{ 6}$ & ( = & 4.89898 & ) \\
  31 &  $(         \pm  4,          \pm  3,          \pm  2)_{\texttt{FS}}$ &  48 & $  \sqrt{29}$ & ( = & 5.38516 & ) \\
  32 &  $(         \pm  4,          \pm  4,          \pm  2)_{\texttt{FS}}$ &  24 & $          6$ & ( = & 6.      & ) \\
  33 &  $(         \pm  4,          \pm  3,          \pm  3)_{\texttt{FS}}$ &  24 & $  \sqrt{34}$ & ( = & 5.83095 & ) \\
  34 &  $(         \pm  4,          \pm  4,          \pm  3)_{\texttt{FS}}$ &  24 & $  \sqrt{41}$ & ( = & 6.40312 & ) \\
  35 &  $(         \pm  4,          \pm  4,          \pm  4)_{\texttt{FS}}$ &   8 & $4 \sqrt{ 3}$ & ( = & 6.9282  & ) \\
  36 &  $(         \pm  5, \phantom{\pm} 0, \phantom{\pm} 0)_{\texttt{FS}}$ &   6 & $          5$ & ( = & 5.      & ) \\
  37 &  $(         \pm  5,          \pm  1, \phantom{\pm} 0)_{\texttt{FS}}$ &  24 & $  \sqrt{26}$ & ( = & 5.09902 & ) \\
  38 &  $(         \pm  5,          \pm  2, \phantom{\pm} 0)_{\texttt{FS}}$ &  24 & $  \sqrt{29}$ & ( = & 5.38516 & ) \\
  39 &  $(         \pm  5,          \pm  3, \phantom{\pm} 0)_{\texttt{FS}}$ &  24 & $  \sqrt{34}$ & ( = & 5.83095 & ) \\
  40 &  $(         \pm  5,          \pm  4, \phantom{\pm} 0)_{\texttt{FS}}$ &  24 & $  \sqrt{41}$ & ( = & 6.40312 & ) \\
  41 &  $(         \pm  5,          \pm  5, \phantom{\pm} 0)_{\texttt{FS}}$ &  12 & $5 \sqrt{ 2}$ & ( = & 7.07107 & ) \\
  42 &  $(         \pm  5,          \pm  1,          \pm  1)_{\texttt{FS}}$ &  24 & $3 \sqrt{ 3}$ & ( = & 5.19615 & ) \\
  \hline
  \end{tabular}
  \begin{tabular}[t]{|c|c|c|rllr|}
  \hline
  Group & Vectors & \# Vectors & \multicolumn{4}{c|}{Length}\\
  \hline
  \hline
  43 &  $(         \pm  5,          \pm  2,          \pm  1)_{\texttt{FS}}$ &  48 & $  \sqrt{30}$ & ( = & 5.47723 & ) \\
  44 &  $(         \pm  5,          \pm  3,          \pm  1)_{\texttt{FS}}$ &  48 & $  \sqrt{35}$ & ( = & 5.91608 & ) \\
  45 &  $(         \pm  5,          \pm  4,          \pm  1)_{\texttt{FS}}$ &  48 & $  \sqrt{42}$ & ( = & 6.48074 & ) \\
  46 &  $(         \pm  5,          \pm  5,          \pm  1)_{\texttt{FS}}$ &  24 & $  \sqrt{51}$ & ( = & 7.14143 & ) \\
  47 &  $(         \pm  5,          \pm  2,          \pm  2)_{\texttt{FS}}$ &  24 & $  \sqrt{33}$ & ( = & 5.74456 & ) \\
  48 &  $(         \pm  5,          \pm  3,          \pm  2)_{\texttt{FS}}$ &  48 & $  \sqrt{38}$ & ( = & 6.16441 & ) \\
  49 &  $(         \pm  5,          \pm  4,          \pm  2)_{\texttt{FS}}$ &  48 & $3 \sqrt{ 5}$ & ( = & 6.7082  & ) \\
  50 &  $(         \pm  5,          \pm  5,          \pm  2)_{\texttt{FS}}$ &  24 & $3 \sqrt{ 6}$ & ( = & 7.34847 & ) \\
  51 &  $(         \pm  5,          \pm  3,          \pm  3)_{\texttt{FS}}$ &  24 & $  \sqrt{43}$ & ( = & 6.55744 & ) \\
  52 &  $(         \pm  5,          \pm  4,          \pm  3)_{\texttt{FS}}$ &  48 & $5 \sqrt{ 2}$ & ( = & 7.07107 & ) \\
  53 &  $(         \pm  5,          \pm  5,          \pm  3)_{\texttt{FS}}$ &  24 & $  \sqrt{59}$ & ( = & 7.68115 & ) \\
  54 &  $(         \pm  5,          \pm  4,          \pm  4)_{\texttt{FS}}$ &  24 & $  \sqrt{57}$ & ( = & 7.54983 & ) \\
  55 &  $(         \pm  5,          \pm  5,          \pm  4)_{\texttt{FS}}$ &  24 & $  \sqrt{66}$ & ( = & 8.12404 & ) \\
  56 &  $(         \pm  5,          \pm  5,          \pm  5)_{\texttt{FS}}$ &   8 & $5 \sqrt{ 3}$ & ( = & 8.66025 & ) \\
  57 &  $(         \pm  6, \phantom{\pm} 0, \phantom{\pm} 0)_{\texttt{FS}}$ &   6 & $          6$ & ( = & 6.      & ) \\
  58 &  $(         \pm  6,          \pm  1, \phantom{\pm} 0)_{\texttt{FS}}$ &  24 & $  \sqrt{37}$ & ( = & 6.08276 & ) \\
  59 &  $(         \pm  6,          \pm  2, \phantom{\pm} 0)_{\texttt{FS}}$ &  24 & $2 \sqrt{10}$ & ( = & 6.32456 & ) \\
  60 &  $(         \pm  6,          \pm  3, \phantom{\pm} 0)_{\texttt{FS}}$ &  24 & $3 \sqrt{ 5}$ & ( = & 6.7082  & ) \\
  61 &  $(         \pm  6,          \pm  4, \phantom{\pm} 0)_{\texttt{FS}}$ &  24 & $2 \sqrt{13}$ & ( = & 7.2111  & ) \\
  62 &  $(         \pm  6,          \pm  5, \phantom{\pm} 0)_{\texttt{FS}}$ &  24 & $  \sqrt{61}$ & ( = & 7.81025 & ) \\
  63 &  $(         \pm  6,          \pm  6, \phantom{\pm} 0)_{\texttt{FS}}$ &  12 & $6 \sqrt{ 2}$ & ( = & 8.48528 & ) \\
  64 &  $(         \pm  6,          \pm  1,          \pm  1)_{\texttt{FS}}$ &  24 & $  \sqrt{38}$ & ( = & 6.16441 & ) \\
  65 &  $(         \pm  6,          \pm  2,          \pm  1)_{\texttt{FS}}$ &  48 & $  \sqrt{41}$ & ( = & 6.40312 & ) \\
  66 &  $(         \pm  6,          \pm  3,          \pm  1)_{\texttt{FS}}$ &  48 & $  \sqrt{46}$ & ( = & 6.78233 & ) \\
  67 &  $(         \pm  6,          \pm  4,          \pm  1)_{\texttt{FS}}$ &  48 & $  \sqrt{53}$ & ( = & 7.28011 & ) \\
  68 &  $(         \pm  6,          \pm  5,          \pm  1)_{\texttt{FS}}$ &  48 & $  \sqrt{62}$ & ( = & 7.87401 & ) \\
  69 &  $(         \pm  6,          \pm  6,          \pm  1)_{\texttt{FS}}$ &  24 & $  \sqrt{73}$ & ( = & 8.544   & ) \\
  70 &  $(         \pm  6,          \pm  2,          \pm  2)_{\texttt{FS}}$ &  24 & $2 \sqrt{11}$ & ( = & 6.63325 & ) \\
  71 &  $(         \pm  6,          \pm  3,          \pm  2)_{\texttt{FS}}$ &  48 & $          7$ & ( = & 7.      & ) \\
  72 &  $(         \pm  6,          \pm  4,          \pm  2)_{\texttt{FS}}$ &  48 & $2 \sqrt{14}$ & ( = & 7.48331 & ) \\
  73 &  $(         \pm  6,          \pm  5,          \pm  2)_{\texttt{FS}}$ &  48 & $  \sqrt{65}$ & ( = & 8.06226 & ) \\
  74 &  $(         \pm  6,          \pm  6,          \pm  2)_{\texttt{FS}}$ &  24 & $2 \sqrt{19}$ & ( = & 8.7178  & ) \\
  75 &  $(         \pm  6,          \pm  3,          \pm  3)_{\texttt{FS}}$ &  24 & $3 \sqrt{ 6}$ & ( = & 7.34847 & ) \\
  76 &  $(         \pm  6,          \pm  4,          \pm  3)_{\texttt{FS}}$ &  48 & $  \sqrt{61}$ & ( = & 7.81025 & ) \\
  77 &  $(         \pm  6,          \pm  5,          \pm  3)_{\texttt{FS}}$ &  48 & $  \sqrt{70}$ & ( = & 8.3666  & ) \\
  78 &  $(         \pm  6,          \pm  6,          \pm  3)_{\texttt{FS}}$ &  24 & $          9$ & ( = & 9.      & ) \\
  79 &  $(         \pm  6,          \pm  4,          \pm  4)_{\texttt{FS}}$ &  24 & $2 \sqrt{17}$ & ( = & 8.24621 & ) \\
  80 &  $(         \pm  6,          \pm  5,          \pm  4)_{\texttt{FS}}$ &  48 & $  \sqrt{77}$ & ( = & 8.77496 & ) \\
  81 &  $(         \pm  6,          \pm  6,          \pm  4)_{\texttt{FS}}$ &  24 & $2 \sqrt{22}$ & ( = & 9.38083 & ) \\
  82 &  $(         \pm  6,          \pm  5,          \pm  5)_{\texttt{FS}}$ &  24 & $  \sqrt{86}$ & ( = & 9.27362 & ) \\
  83 &  $(         \pm  6,          \pm  6,          \pm  5)_{\texttt{FS}}$ &  24 & $  \sqrt{97}$ & ( = & 9.84886 & ) \\
  84 &  $(         \pm  6,          \pm  6,          \pm  6)_{\texttt{FS}}$ &   8 & $6 \sqrt{ 3}$ & ( = & 10.3923 & ) \\
  \hline
  \end{tabular}  
}\caption{Groups of velocity vectors used to generate three-dimensional Cartesian lattices. 
          For each group, we give the vectors forming the set (FS stands for full-symmetric), 
          the cardinality of the group and the length of the vectors belonging to the group. }\label{tab:stencil-groups}
\end{table}
\end{center}

\appendix
\renewcommand{\thesection}{D}
\section{Supplemental Material}\label{sec:appendixD}

The supplemental material, located at \cite{supplementary-material},
includes in a \textit{Mathematica} file full expressions for the following:
\begin{itemize}
  \item Orthogonal relativistic polynomials up to the third order.
  \item Orthogonal projections up to the third order.
  \item Discrete expansion of the equilibrium distribution function.
  \item Examples of quadratures at the second and third order.
\end{itemize}

\section*{References}
\twocolumngrid
\bibliographystyle{elsarticle/elsarticle-num}
\bibliography{biblio}

\begin{thebibliography}{10}
\expandafter\ifx\csname url\endcsname\relax
  \def\url#1{\texttt{#1}}\fi
\expandafter\ifx\csname urlprefix\endcsname\relax\def\urlprefix{URL }\fi
\expandafter\ifx\csname href\endcsname\relax
  \def\href#1#2{#2} \def\path#1{#1}\fi

\bibitem{shore-1992}
S.~N. Shore, \href{http://cds.cern.ch/record/1616948}{{An introduction to
  astrophysical hydrodynamics}}, Elsevier, Burlington, MA, 1992.
\newline\urlprefix\url{http://cds.cern.ch/record/1616948}

\bibitem{muller-2008}
M.~M\"uller, S.~Sachdev, Collective cyclotron motion of the relativistic plasma
  in graphene, Phys. Rev. B 78 (2008) 115419.
\newblock \href {http://dx.doi.org/10.1103/PhysRevB.78.115419}
  {\path{doi:10.1103/PhysRevB.78.115419}}.

\bibitem{wan-2011}
X.~Wan, A.~M. Turner, A.~Vishwanath, S.~Y. Savrasov, Topological semimetal and
  fermi-arc surface states in the electronic structure of pyrochlore iridates,
  Phys. Rev. B 83 (2011) 205101.
\newblock \href {http://dx.doi.org/10.1103/PhysRevB.83.205101}
  {\path{doi:10.1103/PhysRevB.83.205101}}.

\bibitem{ackermann-2001}
K.~H.~A. et~al., Elliptic flow in $au+au$ collisions at
  $\ensuremath{\surd}{s}_{\mathrm{nn}}\phantom{\rule{0ex}{0ex}}=\phantom{\rule{0ex}{0ex}}130\mathrm{GeV}$,
  Phys. Rev. Lett. 86 (2001) 402--407.
\newblock \href {http://dx.doi.org/10.1103/PhysRevLett.86.402}
  {\path{doi:10.1103/PhysRevLett.86.402}}.

\bibitem{succi-2001}
S.~Succi, The {Lattice Boltzmann Equation} for Fluid Dynamics and Beyond,
  Clarendon Press, Oxford, 2001.
\newblock \href {http://dx.doi.org/10.1016/S0997-7546(02)00005-5}
  {\path{doi:10.1016/S0997-7546(02)00005-5}}.

\bibitem{aidun-2010}
C.~K. Aidun, J.~R. Clausen, Lattice-boltzmann method for complex flows, Annual
  Review of Fluid Mechanics 42~(1) (2010) 439--472.
\newblock \href {http://dx.doi.org/10.1146/annurev-fluid-121108-145519}
  {\path{doi:10.1146/annurev-fluid-121108-145519}}.

\bibitem{succi-2015}
{Succi, Sauro}, Lattice boltzmann 2038, EPL 109~(5) (2015) 50001.
\newblock \href {http://dx.doi.org/10.1209/0295-5075/109/50001}
  {\path{doi:10.1209/0295-5075/109/50001}}.

\bibitem{tolke-2008}
J.~T\"olke, M.~Krafczyk, {TeraFLOP computing on a desktop PC with GPUs for 3D
  CFD}, International Journal of Computational Fluid Dynamics 22~(7) (2008)
  443--456.
\newblock \href {http://dx.doi.org/10.1080/10618560802238275}
  {\path{doi:10.1080/10618560802238275}}.

\bibitem{bernaschi-2009}
M.~Bernaschi, S.~Melchionna, S.~Succi, M.~Fyta, E.~Kaxiras, J.~Sircar, Muphy: A
  parallel {MU}lti {PHY}sics/scale code for high performance bio-fluidic
  simulations, Computer Physics Communications 180~(9) (2009) 1495 -- 1502.
\newblock \href {http://dx.doi.org/10.1016/j.cpc.2009.04.001}
  {\path{doi:10.1016/j.cpc.2009.04.001}}.

\bibitem{mountrakis-2015}
L.~Mountrakis, E.~Lorenz, O.~Malaspinas, S.~Alowayyed, B.~Chopard, A.~G.
  Hoekstra, Parallel performance of an ib-lbm suspension simulation framework,
  Journal of Computational Science 9 (2015) 45 -- 50.
\newblock \href {http://dx.doi.org/10.1016/j.jocs.2015.04.006}
  {\path{doi:10.1016/j.jocs.2015.04.006}}.

\bibitem{calore-2016}
E.~Calore, A.~Gabbana, J.~Kraus, E.~Pellegrini, S.~F. Schifano, R.~Tripiccione,
  Massively parallel lattice-boltzmann codes on large {GPU} clusters, Parallel
  Computing 58 (2016) 1--24.
\newblock \href {http://dx.doi.org/10.1016/j.parco.2016.08.005}
  {\path{doi:10.1016/j.parco.2016.08.005}}.

\bibitem{mendoza-2010a}
M.~Mendoza, B.~M. Boghosian, H.~J. Herrmann, S.~Succi, Fast lattice boltzmann
  solver for relativistic hydrodynamics, Phys. Rev. Lett. 105 (2010) 014502.
\newblock \href {http://dx.doi.org/10.1103/PhysRevLett.105.014502}
  {\path{doi:10.1103/PhysRevLett.105.014502}}.

\bibitem{mendoza-2010b}
M.~Mendoza, B.~M. Boghosian, H.~J. Herrmann, S.~Succi, Derivation of the
  lattice boltzmann model for relativistic hydrodynamics, Phys. Rev. D 82
  (2010) 105008.
\newblock \href {http://dx.doi.org/10.1103/PhysRevD.82.105008}
  {\path{doi:10.1103/PhysRevD.82.105008}}.

\bibitem{romatschke-2011}
P.~Romatschke, M.~Mendoza, S.~Succi, Fully relativistic lattice boltzmann
  algorithm, Phys. Rev. C 84 (2011) 034903.
\newblock \href {http://dx.doi.org/10.1103/PhysRevC.84.034903}
  {\path{doi:10.1103/PhysRevC.84.034903}}.

\bibitem{li-2012}
Q.~Li, K.~H. Luo, X.~J. Li, {Lattice Boltzmann method for relativistic
  hydrodynamics: Issues on conservation law of particle number and
  discontinuities}, Phys. Rev. D86 (2012) 085044.
\newblock \href {http://dx.doi.org/10.1103/PhysRevD.86.085044}
  {\path{doi:10.1103/PhysRevD.86.085044}}.

\bibitem{mohseni-2013}
F.~Mohseni, M.~Mendoza, S.~Succi, H.~J. Herrmann, Lattice boltzmann model for
  ultrarelativistic flows, Phys. Rev. D 87 (2013) 083003.
\newblock \href {http://dx.doi.org/10.1103/PhysRevD.87.083003}
  {\path{doi:10.1103/PhysRevD.87.083003}}.

\bibitem{mendoza-2013}
M.~Mendoza, I.~Karlin, S.~Succi, H.~J. Herrmann, Relativistic lattice boltzmann
  model with improved dissipation, Phys. Rev. D 87 (2013) 065027.
\newblock \href {http://dx.doi.org/10.1103/PhysRevD.87.065027}
  {\path{doi:10.1103/PhysRevD.87.065027}}.

\bibitem{supplementary-material}
\href{http://link.aps.org/supplemental/10.1103/PhysRevE.95.053304}{See
  supplemental material at}.
\newline\urlprefix\url{http://link.aps.org/supplemental/10.1103/PhysRevE.95.053304}

\bibitem{bhatnagar-1954}
P.~L. Bhatnagar, E.~P. Gross, M.~Krook, A model for collision processes in
  gases. amplitude processes in charged and neutral one-component systems,
  Phys. Rev. 94~(3) (1954) 511--525.
\newblock \href {http://dx.doi.org/10.1103/PhysRev.94.511}
  {\path{doi:10.1103/PhysRev.94.511}}.

\bibitem{anderson-witting-1974a}
J.~Anderson, H.~Witting, A relativistic relaxation-time model for the boltzmann
  equation, Physica 74~(3) (1974) 466 -- 488.
\newblock \href {http://dx.doi.org/10.1016/0031-8914(74)90355-3}
  {\path{doi:10.1016/0031-8914(74)90355-3}}.

\bibitem{anderson-witting-1974b}
J.~Anderson, H.~Witting, Relativistic quantum transport coefficients, Physica
  74~(3) (1974) 489 -- 495.
\newblock \href {http://dx.doi.org/10.1016/0031-8914(74)90356-5}
  {\path{doi:10.1016/0031-8914(74)90356-5}}.

\bibitem{grad-1949}
H.~Grad, On the kinetic theory of rarefied gases, Communications on Pure and
  Applied Mathematics 2~(4) (1949) 331--407.
\newblock \href {http://dx.doi.org/10.1002/cpa.3160020403}
  {\path{doi:10.1002/cpa.3160020403}}.

\bibitem{cercignani-2002}
C.~Cercignani, G.~M. Kremer, The Relativistic Boltzmann Equation: Theory and
  Applications, Birkhäuser Basel, 2002.
\newblock \href {http://dx.doi.org/10.1007/978-3-0348-8165-4}
  {\path{doi:10.1007/978-3-0348-8165-4}}.

\bibitem{higuera-1989}
F.~J. Higuera, S.~Succi, R.~Benzi,
  \href{http://stacks.iop.org/0295-5075/9/i=4/a=008}{Lattice gas dynamics with
  enhanced collisions}, EPL (Europhysics Letters) 9~(4) (1989) 345.
\newline\urlprefix\url{http://stacks.iop.org/0295-5075/9/i=4/a=008}

\bibitem{xiaoyi-luo-1997}
X.~He, L.-S. Luo, Theory of the lattice boltzmann method: From the boltzmann
  equation to the lattice boltzmann equation, Phys. Rev. E 56 (1997)
  6811--6817.
\newblock \href {http://dx.doi.org/10.1103/PhysRevE.56.6811}
  {\path{doi:10.1103/PhysRevE.56.6811}}.

\bibitem{shan-1997}
X.~{Shan}, X.~{He}, {Discretization of the velocity space in solution of the
  Boltzmann equation}\href {http://dx.doi.org/10.1103/PhysRevLett.80.65}
  {\path{doi:10.1103/PhysRevLett.80.65}}.

\bibitem{martys-1998}
N.~S. Martys, X.~Shan, H.~Chen, Evaluation of the external force term in the
  discrete boltzmann equation, Phys. Rev. E 58 (1998) 6855--6857.
\newblock \href {http://dx.doi.org/10.1103/PhysRevE.58.6855}
  {\path{doi:10.1103/PhysRevE.58.6855}}.

\bibitem{karsch-1980}
F.~Karsch, D.~E. Miller, Exact equation of state for ideal relativistic quantum
  gases, Phys. Rev. A 22 (1980) 1210--1219.
\newblock \href {http://dx.doi.org/10.1103/PhysRevA.22.1210}
  {\path{doi:10.1103/PhysRevA.22.1210}}.

\bibitem{israel-1976}
W.~Israel, J.~Stewart, Thermodynamics of nonstationary and transient effects in
  a relativistic gas, Physics Letters A 58~(4) (1976) 213 -- 215.
\newblock \href {http://dx.doi.org/10.1016/0375-9601(76)90075-X}
  {\path{doi:10.1016/0375-9601(76)90075-X}}.

\bibitem{denicol-2012}
G.~S. Denicol, H.~Niemi, E.~Moln\'ar, D.~H. Rischke, Derivation of transient
  relativistic fluid dynamics from the boltzmann equation, Phys. Rev. D 85
  (2012) 114047.
\newblock \href {http://dx.doi.org/10.1103/PhysRevD.85.114047}
  {\path{doi:10.1103/PhysRevD.85.114047}}.

\bibitem{philippi-2006}
P.~C. Philippi, L.~A. Hegele, L.~O.~E. dos Santos, R.~Surmas, From the
  continuous to the lattice boltzmann equation: The discretization problem and
  thermal models, Phys. Rev. E 73 (2006) 056702.
\newblock \href {http://dx.doi.org/10.1103/PhysRevE.73.056702}
  {\path{doi:10.1103/PhysRevE.73.056702}}.

\bibitem{shan-2010}
X.~Shan, General solution of lattices for cartesian lattice
  bhatanagar-gross-krook models, Phys. Rev. E 81 (2010) 036702.
\newblock \href {http://dx.doi.org/10.1103/PhysRevE.81.036702}
  {\path{doi:10.1103/PhysRevE.81.036702}}.

\bibitem{rolando-2014}
V.~Rolando, G.~Inghirami, A.~Beraudo, L.~D. Zanna, F.~Becattini, V.~Chandra,
  A.~D. Pace, M.~Nardi, Heavy ion collision evolution modeling with echo-qgp,
  Nuclear Physics A 931 (2014) 970 -- 974.
\newblock \href {http://dx.doi.org/10.1016/j.nuclphysa.2014.08.011}
  {\path{doi:10.1016/j.nuclphysa.2014.08.011}}.

\bibitem{xu-greiner-2007}
Z.~Xu, C.~Greiner, Transport rates and momentum isotropization of gluon matter
  in ultrarelativistic heavy-ion collisions, Phys. Rev. C 76 (2007) 024911.
\newblock \href {http://dx.doi.org/10.1103/PhysRevC.76.024911}
  {\path{doi:10.1103/PhysRevC.76.024911}}.

\bibitem{molnar-2014}
E.~Moln\'ar, H.~Niemi, G.~S. Denicol, D.~H. Rischke, Relative importance of
  second-order terms in relativistic dissipative fluid dynamics, Phys. Rev. D
  89 (2014) 074010.
\newblock \href {http://dx.doi.org/10.1103/PhysRevD.89.074010}
  {\path{doi:10.1103/PhysRevD.89.074010}}.

\bibitem{taylor-1953}
G.~Taylor, Dispersion of soluble matter in solvent flowing slowly through a
  tube, Proceedings of the Royal Society of London A: Mathematical, Physical
  and Engineering Sciences 219~(1137) (1953) 186--203.
\newblock \href {http://dx.doi.org/10.1098/rspa.1953.0139}
  {\path{doi:10.1098/rspa.1953.0139}}.

\bibitem{cooper-1974}
F.~Cooper, G.~Frye, Single-particle distribution in the hydrodynamic and
  statistical thermodynamic models of multiparticle production, Phys. Rev. D 10
  (1974) 186--189.
\newblock \href {http://dx.doi.org/10.1103/PhysRevD.10.186}
  {\path{doi:10.1103/PhysRevD.10.186}}.

\bibitem{heinz-2005}
U.~W. Heinz, {'RHIC serves the perfect fluid': Hydrodynamic flow of the QGP},
  in: {Proceedings, Workshop on Extreme QCD, 2005}, 2005, pp. 3--12.
\newblock \href {http://arxiv.org/abs/nucl-th/0512051}
  {\path{arXiv:nucl-th/0512051}}.

\bibitem{chojnacki-2005}
M.~Chojnacki, W.~Florkowski, T.~Cs\"org\"o,
  \href{http://link.aps.org/doi/10.1103/PhysRevC.71.044902}{Formation of
  hubble-like flow in little bangs}, Phys. Rev. C 71 (2005) 044902.
\newblock \href {http://dx.doi.org/10.1103/PhysRevC.71.044902}
  {\path{doi:10.1103/PhysRevC.71.044902}}.
\newline\urlprefix\url{http://link.aps.org/doi/10.1103/PhysRevC.71.044902}

\end{thebibliography}


\end{document}